\pdfoutput=1
\documentclass[twocolumn]{svjour3}          
\smartqed  
\smartqed  
\usepackage{graphicx}
\usepackage{mathptmx}      
\usepackage{latexsym}
\usepackage{ifpdf}
\usepackage{balance}
\usepackage{graphicx}
\usepackage{multirow}
\usepackage{adjustbox}
\usepackage{wrapfig}
\usepackage{subfig}
\usepackage[colorlinks=true,allcolors=blue]{hyperref}%
\usepackage[linesnumbered,ruled,vlined]{algorithm2e}
\usepackage{comment}
\graphicspath{{./figure/}}

\usepackage{hyphenat}
\UseRawInputEncoding

\usepackage[misc,geometry]{ifsym}
\newcommand{\compact}{\vspace{-0pt}}
\newcommand{\subcompact}{\vspace{-0pt}}

\usepackage[skip=5pt]{caption}
\usepackage{cite}
\usepackage{lipsum}  

\newcommand{\change}[1]{{\textcolor{black}{#1}}}

\newcommand{\acid}{\texttt{ACID}\xspace}

\newcommand{\tsp}{{TSP}\xspace}

\usepackage{array,booktabs}
\newcolumntype{P}[1]{>{\centering\arraybackslash}p{#1}}

\usepackage{array}

\newcolumntype{L}[1]{>{\raggedright\let\newline\\\arraybackslash\hspace{0pt}}m{#1}}
\newcolumntype{C}[1]{>{\centering\let\newline\\\arraybackslash\hspace{0pt}}m{#1}}
\newcolumntype{R}[1]{>{\raggedleft\let\newline\\\arraybackslash\hspace{0pt}}m{#1}}

\usepackage{tcolorbox}
\usepackage{marginnote}
\newcommand{\margi}[1]{
\marginnote{
}
}

\usepackage{float}

\begin{document}

\title{A Survey on Transactional Stream Processing}
 
\author{
	Shuhao Zhang
	\and
        Juan Soto
        \and
        Volker Markl 
}

\institute{\Letter\  Shuhao Zhang \at
              Singapore University of Technology and Design, Singapore \\
              \email{shuhao\_zhang@sutd.edu.sg}           
           \and
           Juan Soto \at
              Technische Universit{\"a}t Berlin, Germany \\
              \email{juan.soto@tu-berlin.de}           
           \and              
           Volker Markl \at
              Technische Universit{\"a}t Berlin, Germany \\
              \email{volker.markl@tu-berlin.de}           
}

\date{Received: date / Accepted: date}

\maketitle

\compact
\begin{abstract}
Transactional stream processing (TSP) strives to create a cohesive model that merges the advantages of both transactional and stream-oriented guarantees. Over the past decade, numerous endeavors have contributed to the evolution of TSP solutions, uncovering similarities and distinctions among them. Despite these advances, a universally accepted standard approach for integrating transactional functionality with stream processing remains to be established. Existing TSP solutions predominantly concentrate on specific application characteristics and involve complex design trade-offs. This survey intends to introduce TSP and present our perspective on its future progression. Our primary goals are twofold: to provide insights into the diverse TSP requirements and methodologies, and to inspire the design and development of groundbreaking TSP systems.
 
\keywords{Transactions \and Stream Processing \and Survey}
\end{abstract}
\compact
\section{Introduction}
\label{sec:intro}
\begin{figure}[t]
\centering	
	\includegraphics*[width=0.4\textwidth]{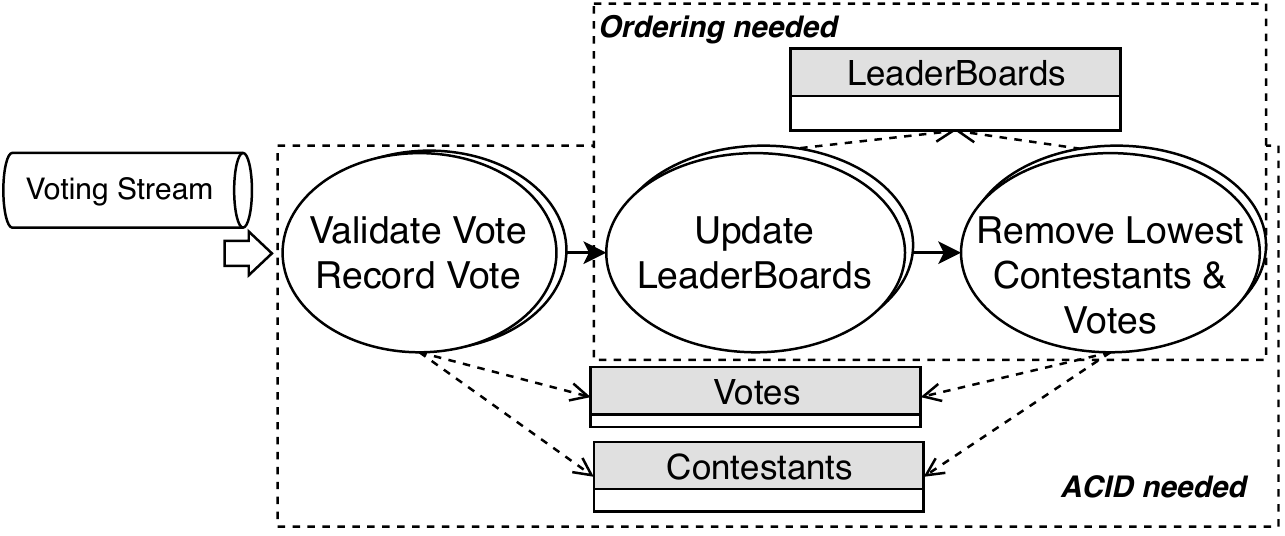}
	\caption{Leaderboard Maintenance (\textsl{LM})~\cite{S-Store}.}
	\label{fig:lm} 
\end{figure} 
Stream processing has emerged as a powerful paradigm for real-time data processing, with its roots dating back to the late 1990s and early 2000s when publish-subscribe systems and data stream management systems (DSMS) were introduced~\cite{babcock2002models}. Over time, various stream processing frameworks, such as Apache Storm, Apache Flink, and Apache Kafka Streams, have been developed, each with its own strengths and weaknesses. 

The concept of combining relational queries with continuous stream processing has been proposed since the inception of the first-generation stream processing engine (SPE)~\cite{golab1}. 
\change{However, most modern SPEs either disallow the maintenance of relational tables~\cite{10.1145/1007568.1007616} or permit (shared) relational tables to be queried during continuous query processing, but implicitly assume that relations remain unchanged throughout the lifespan of\margi{R1Q11, R1Q14}a query~\cite{Aurora,madden2002fjording}.
Consequently, current SPEs have some limitations. 
First, they do not offer transactional guarantees. 
Second, they are unable to maintain a consistent state across distributed systems, such as the management of transactional records during stream processing~\cite{S-Store}.} 
This has led to the emergence of TSP systems~\cite{conway2008cisc,Transactions2018,AFFETTI202065,Arasu:2006:CCQ:1146461.1146463,botan2012transactional,meehan2017data,stream2003stream}. 

\change{In\margi{R1Q2}general, a TSP system is a computing system that processes continuous streams of data while providing transactional guarantees, such as atomicity, consistency, isolation, and durability (ACID)~\cite{Tatbul2018}. To qualify as a TSP system, a system should possess the following two fundamental characteristics: 1) \textit{enable continuous data processing}: the system should be able to process incoming data streams in (near)real-time, without the need for batch processing or storing the entire dataset in memory; 2) \textit{offer transactional guarantees}: the system should guarantee ACID properties are met, to ensure that data operations are correct, reliable, and maintain data integrity. These systems are designed to handle real-time data processing and analytics, while ensuring the correctness and integrity of data operations.
}
\subcompact
\subsection{Example Use Cases}
To provide context, next, we discuss two\margi{R1Q8, R3Q1}representative scenarios as motivating examples, which is then followed by some remarks. A detailed list of relevant application scenarios is presented in Section~\ref{sec:app}.

\textbf{Leaderboard Maintenance.}
Meehan et al.~\cite{S-Store} describe a leaderboard maintenance application (depicted in Figure~\ref{fig:lm}), which considers a TV show akin to American Idol. In this use case, viewers vote for their favorite contestants (via text message), and the system must continuously and accurately update and display the top contestants ranked by the total number of votes in real-time. Suppose a preferential voting scheme is employed, where the contestant with the fewest votes is periodically eliminated. Additionally, several leaderboards are maintained: one for the top-3 contestants, one for the bottom-3 contestants, and a third one for the top-3 trending contestants over the last 100 votes. With each incoming vote, these three leaderboards are continuously updated. Here, there are two key requirements. First, incoming votes need to be validated and recorded in a shared table called \textit{Votes}, which requires ACID guarantees. Second, to ensure fairness, early votes need to be tallied before later votes, thereby necessitating an ordering guarantee.

\begin{figure}[t]
\centering	
	\includegraphics*[width=0.4\textwidth]{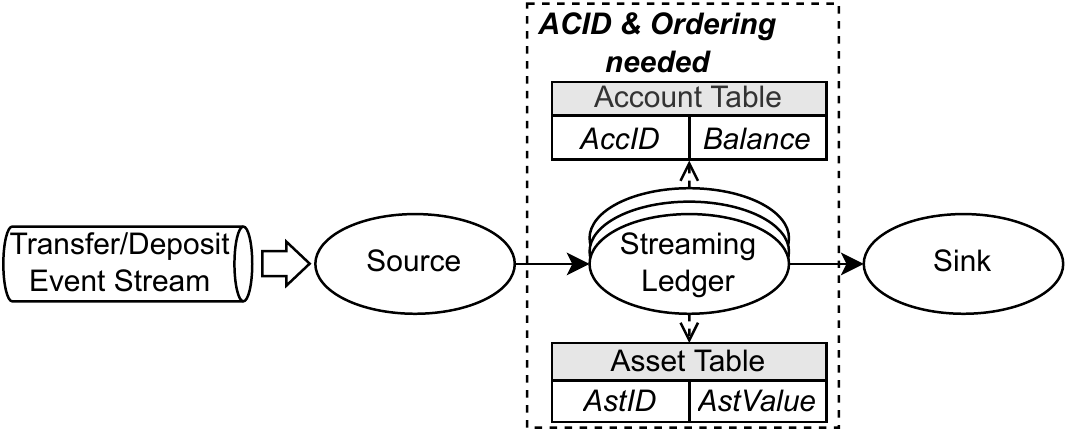}
	\caption{Streaming Ledger (SL)~\cite{Transactions2018}.}
	\label{fig:SL_exmaple} 
\end{figure}
\textbf{Streaming Ledger.}
Another simplified use case that benefits from TSP is the Streaming Ledger (SL), proposed by dataArtisans~\cite{Transactions2018}, as illustrated in Figure~\ref{fig:SL_exmaple}. In this scenario, two types of requests access two tables with shared mutable states: \textit{transfers}, which exchange funds between user accounts or assets, and \textit{deposits}, which increase the balance on user accounts or assets. The aim is to process a stream of these requests and output the results. To handle a large volume of concurrent requests both ACID guarantees and ordering guarantees are necessary. For instance, transfers may be rolled back, if they violate data integrity, such as causing a negative account balance. Additionally, the system may need to process requests in temporal order by timestamp.

\textbf{Remarks.}
Neither modern SPEs, nor databases adequately support the two aforementioned use cases. On the one hand, modern SPEs restrict each execution entity (e.g., each operator/thread), to maintain a disjoint subset of states (or partitioned states)~\cite{flink}. However, this violates a transactional consistency guarantee, when each input tuple involves multiple keys. On the other hand, databases were not designed to handle workloads with high-velocity input stream insertion. Contrastingly, TSP offers both ACID and stream processing properties in a unified manner.

Integrating streaming and transactional features is a complex task. Over the past decade, various research groups have developed their own definitions and corresponding solutions for TSP~\cite{golab2,Chen10,DSPCC,TSDM,botan2012transactional,tstream,Affetti:2017:FIS:3093742.3093929}. 
This diversity stems from the wide range of application scenarios proposed in the literature, as well as the fact that each system concentrates on a specific use case, which imposes implicit assumptions and objectives. Use cases that benefit from TSP typically employ streaming capabilities to persist state or provide (near real-time) views of shared tables, while simultaneously leveraging transactional features, to ensure a consistent representation of the state or a summary of the shared tables. 

\subcompact
\subsection{Scope}
This survey provides\margi{R1Q10} a comprehensive overview of TSP, encompassing its background, properties, design aspects, technologies, systems, and applications. We delve into the challenges and design trade-offs in implementing TSP and provide insights into the current state of the field and potential future directions. Our analysis focuses primarily on stateful stream processing, which involves the general task of processing dynamic or streaming data with algorithms that can continuously produce new results on-the-fly. While we discuss various aspects of stream processing, including dataflow engines and distributed systems, our analysis is not limited to these specific design and implementation choices. The survey emphasizes the following six aspects:
\begin{itemize}
    \item \textit{Background}: Discusses the history and motivation behind stream processing and TSP, and outlines various transaction models over data streams.
    \item \textit{Properties}: Explores the key properties of TSP, such as transactions, delivery guarantees, and state management.
    \item \textit{Design Aspects}: Examines the design aspects of TSP, including languages, APIs, and system architectures.
    \item \textit{Technologies}: Investigates the technologies used to implement TSP and how each property is addressed with potential alternatives.
    \item \textit{Systems}: Surveys representative TSP systems, focusing on their features, strengths, and weaknesses.
    \item \textit{Applications}: Presents real-world applications and scenarios where TSP is leveraged to provide valuable insights and improved decision-making.
\end{itemize}

\subcompact
\subsection{Outline of the Survey}
\change{The subsequent~\margi{R1Q5}sections are organized as follows. 
Section~\ref{sec:background} introduces the background and overview of TSP, covering the conceptual framework of TSP systems, transaction models over data streams, and providing a high-level overview of the survey.
Section~\ref{sec:taxonomy} discusses the taxonomy of TSP systems, delving into system properties, system architecture and design aspects, and the technologies used to implement TSP.
Section~\ref{sec:systems} presents a survey of TSP systems, including early and recent TSP systems, and compares them based on the taxonomy established in Section~\ref{sec:taxonomy}.
}
Section~\ref{sec:app} showcases various applications and use cases of TSP, such as stream processing optimization, concurrent stateful processing, and stream and DBMS integration.
Section~\ref{sec:future} highlights open challenges and future directions for TSP research and development, including novel applications and  hardware platforms.
Section~\ref{sec:conclusion} concludes with a summary of the key findings and insights gained throughout the survey.
\compact
\section{Background}
\label{sec:background}
In this section, we provide a brief overview of key terms and definitions essential for understanding transactional stream processing systems, followed by a conceptual framework of TSP, and lastly, discuss several transaction models over data streams. The definitions of terms and explanations of concepts are drawn from various sources, including Babcock et al.~\cite{babcock2002models}.

\subcompact
\subsection{Terms and Definitions}
A \textit{data stream} represents one or more underlying signals, such as a network traffic stream indicating the type and volume of data transmitted among nodes in a network. An \textit{event} $e$ is a 2-tuple $e = <t, v>$, consisting of a timestamp and payload. The \textit{timestamp} specifies when an event occurred, while the payload represents data values used during stream processing.

\textit{Stream queries} are collections of \textit{operators} that continuously process streaming events. \change{An \textit{operator} can be\margi{R1Q28}broadly defined as a fundamental component or unit of computation within a stream processing system.} Traditional stream operators include join and aggregation, while modern SPEs also support user-defined stream operators.

\textit{Windows} define bounded subsets of streams to manage infinite streams, with various window types such as tumbling, sliding, and session windows. The \textit{state} is a fundamental concept in stream processing, enabling the comparison of current data with historical data.

\change{A \textit{dataflow model} is a powerful abstraction for stream processing systems,\margi{R1Q12} that enables parallel and distributed processing by breaking tasks into smaller units and orchestrates their execution across multiple nodes. In this model stream processing tasks are represented as directed acyclic graphs (DAGs) with nodes representing operators and edges representing the flow of data between operators. Utilizing a dataflow model allows stream processing systems to distribute workload efficiently, which in turn may increase the demand for transactional correctness guarantees.}


\subcompact
\subsection{Conceptual Framework of TSP}
This section provides\margi{R1Q4, R1Q5, R1Q62} a comprehensive conceptual framework for transactional stream processing (TSP) systems, discussing the relationships between the key components and aspects. A TSP system is comprised of five components: \textit{transactions}, \textit{transaction models}, \textit{operators}, \textit{scheduler}, and \textit{storage}. These components are closely related to the four key aspects of a TSP system, which are language, programming model, execution model, and architecture.

\textit{Transactions} are critical components of TSP systems, ensuring that data is processed and maintained consistently and reliably. Transactions combine the real-time nature of stream processing with the reliability and consistency guarantees of traditional transactional systems. \margi{R1Q37}\textit{Transaction models} describe the granularity and scope of transactions within a TSP system. 
In TSP, a transaction is considered committed when it has successfully completed an operation (e.g., insertion, update, deletion) and the system has confirmed that transactional changes are consistent with the desired transactional guarantees (e.g., consistency, isolation, durability). Upon committing a transaction, the system ensures that its effects are persistent and can be recovered in the event of a failure. This concept of ``commit'' is essential in TSP systems to maintain the integrity and consistency of the data during processing.

\textit{Operators} are responsible for processing incoming and outgoing data, and performing operations, such as filtering, aggregation, transformation, or joins. Operators may also have mutable state that needs to be managed in a transactional manner. \textit{Scheduler} manages the execution of operators and ensures that transactions are executed in the correct order, according to the chosen consistency and isolation models. Schedulers may need to handle out-of-order events, coordinate distributed execution, and manage resource allocation. The \textit{storage} component manages the persistent state of the system, including the mutable state of operators and any intermediate or final results produced during processing. The storage component must also ensure durability and fault tolerance, while providing efficient access to the data.

The conceptual framework of a TSP system covers the key aspects to consider when designing and implementing a TSP system, including language, programming model, execution model, and architecture.

\emph{Language:}
The language aspect of a TSP system defines the syntax and semantics for expressing streaming and on-demand queries, as well as the transactional properties (e.g., consistency, isolation, durability). This aspect is related to the transaction models and operators, as it provides the means for developers to define and manipulate transactions and the operations they perform.

\emph{Programming Model:}
The programming model refers to the way developers interact with the TSP system to define and manage stateful operations and transactional guarantees. This aspect is related to transactions, transaction models, and operators, as the programming model provides the framework for working with these components in a structured and organized manner.

\emph{Execution Model:}
The execution model focuses on how the TSP system processes both streaming and on-demand queries while providing transactional guarantees. This aspect is related to the scheduler and operators, as the execution model determines how the scheduler manages the execution of operators and ensures that transactions are executed in the correct order, according to their consistency and isolation models.

\emph{Architecture:}
The architecture aspect of a TSP system addresses how the system is designed and structured to balance the requirements of stream processing and data management. This aspect is related to all five of the key components. 
The architecture facilitates interaction between these components, such as sharing state between on-demand and streaming queries and providing read and write access as needed.

\subcompact
\subsection{Transaction Models over Data Streams}
Next, we discuss some notable transaction models over data streams, along with their implementation approaches, to provide a comprehensive understanding of transactional guarantees in stream processing.

\subcompact
\subsubsection{Abstract Models}
Various transaction models have been explored for stream processing applications to address consistency guarantees. These models serve as processing paradigms in TSP systems.

\textbf{Per-tuple Transactions:} Each tuple in the data stream is treated as a separate transaction, adhering to ACID properties. This approach is suitable for scenarios requiring atomic and isolated processing of individual events. However, it may introduce significant overhead due to frequent coordination between processing nodes.

\textbf{Micro-batch Transactions:} Data streams are divided into small, bounded micro-batches, with transactions executed over them. This approach reduces the overhead associated with per-tuple transactions, allowing for parallelism and optimization opportunities. However, it may introduce additional latency.

\textbf{Window-based Transactions:} Transactions are executed over windows based on a unit of time, the number of events, or some other criterion. This approach enables processing of related events within the same transaction, thereby, providing stronger consistency guarantees. Yet this can be challenging to manage, especially when dealing with out-of-order events or evolving window types.

\textbf{Group-based Transactions:} Transactions are executed over groups of related events in the data stream. Groups can be defined based on specific criteria, providing fine-grained control over transaction boundaries, and stronger consistency guarantees for complex processing tasks. However, managing group-based transactions can be complex, requiring the identification of related events and coordination across multiple processing nodes.

\textbf{Adaptive Transactions:} This flexible model allows the system to dynamically adjust transaction boundaries based on workload, system state, or application requirements. This can include switching between different transaction models or adapting granularity to optimize performance or meet specific consistency guarantees. Implementing adaptive transactions can be challenging, as it requires real-time monitoring and the adaptation of transaction boundaries.

\subcompact
\subsubsection{Implementation Approaches}
Implementation approaches for the aforementioned transaction models can be classified into three categories: unified transactions, embedded-transactions, and state transactions. Depending on an application's specific requirements, an appropriate combination of implementation approach and particular transaction model should be chosen to achieve the desired performance, scalability, and fault tolerance capability. 

\textbf{Unified Transactions:} 
This approach embeds stream processing operations into a transaction, providing a single framework for handling both stream processing and transactions.
Unified transactions can potentially support various transaction models, as it allows flexibility in defining the scope and granularity of transactions. However, it might be more suitable for fine-grained transaction models, such as per-tuple or micro-batch transactions.

\textbf{Embedded Transactions:} 
This approach embeds transaction processing into stream processing, allowing for transactional semantics without the need for separate transaction management. 
Embedded transactions can be more efficient for certain transaction models, particularly when a lightweight transaction mechanism is required. It might be better suited for per-tuple, micro-batch, or adaptive transactions, where the overhead of separate transaction management can be minimized.

\textbf{State Transactions:} 
This approach separates transaction processing and stream processing, focusing on managing shared mutable state through transactions. 
State transactions can also support different transaction models, but are more suitable for scenarios where state management is a primary concern, such as window-based, group-based, or adaptive transactions.
\compact
\section{Taxonomy of TSP}
\label{sec:taxonomy}
\begin{figure*}[t]
\centering	
	\includegraphics*[width=1\textwidth]{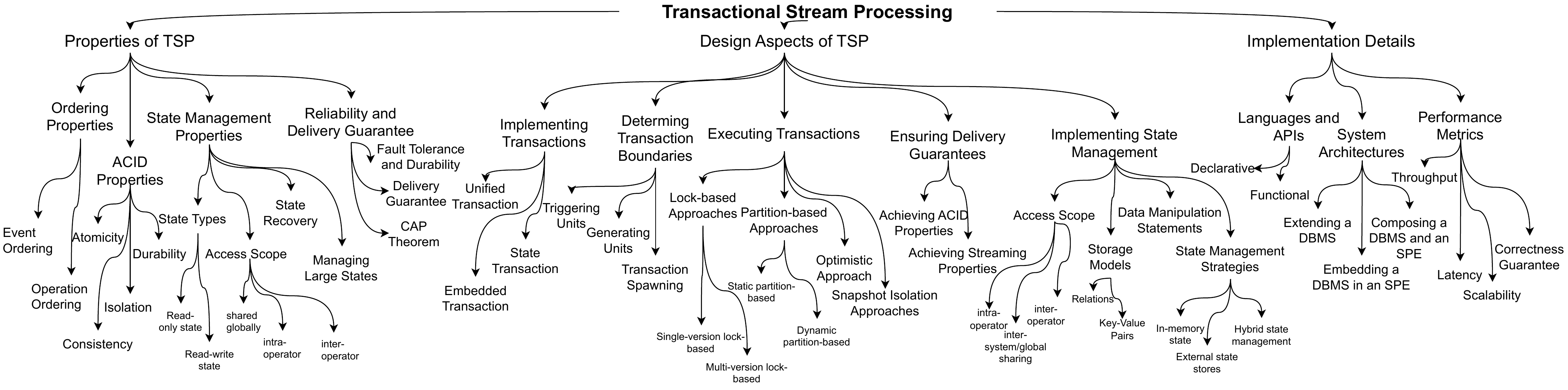}
    \caption{Taxonomy of TSP.}
	\label{fig:taxonomy}
\end{figure*}

\margi{R1Q5, R1Q15}
In this section, we examine 
a taxonomy of Transactional Stream Processing (TSP) as illustrated in  
Figure~\ref{fig:taxonomy}. 
The taxonomy is structured into three key categories: 
\emph{Properties of TSP:} Here we examine the characteristics and requirements of TSP systems, including ordering, ACID properties, state management, and reliability. 
Analyzing these properties enables us to better understand the fundamental issues and challenges prevalent in TSP systems that seek to ensure accurate and reliable data processing. 
\emph{Design Aspects of TSP:} Here we explore the design considerations in TSP systems, spanning transaction implementation, boundaries, execution, delivery guarantees, and state management. Investigating these design aspects aids us in evaluating the suitability of TSP systems for specific applications and provides insights into various design choices and trade-offs. 
\emph{Implementation Details:} Here we address the practical aspects of TSP system implementation, such as programming languages, APIs, system architectures, and component integration. This section also discusses performance metrics and evaluation criteria for TSP systems. By examining these implementation details, we gain a deeper understanding of the practical challenges and proposed solutions in the development of TSP systems. Ultimately, this enables us to be more informed about the choices that must be made when designing or selecting which TSP system to employ for a given use case.

\subsection{Properties of TSP}
\label{sec:properties}
\change{
A TSP model\margi{R3Q7}is required to meet both the ordering properties of streaming operators and events, and the ACID properties of transactions, while also addressing state management, reliability, fault tolerance, and durability. These elements along with the CAP theorem's implications are crucial to the design and functionality of TSP systems. Hence, we will delve into these properties in the subsequent sections, starting with ordering properties followed by ACID properties.
}
\subsubsection{Ordering Properties}
In TSP systems, there are two critical ordering properties: event ordering and operation ordering. 

\subcompact
\paragraph{Event Ordering:}
\change{Event\margi{R1Q42}ordering requires that transactions be processed according to the order of their triggering events, typically based on timestamps or some other logical ordering mechanism. This property is crucial to ensure transactions are executed in the correct sequence, so as to avoid inconsistencies. 
It also helps prevent race conditions or out-of-order processing, which can occur in distributed TSP systems with high levels of parallelism. It is worth noting that the ordering schedule is determined explicitly by the input event rather than the transaction execution order. 
}

\change{
To maintain event order 
TSP systems may employ strategies such as locking, versioning, or optimistic concurrency control. Golab et al.\cite{golab2} propose two stronger serialization properties with ordering guarantees. The first is called \emph{window-serializable}, which requires a read-only transaction to perform a read either strictly before a window is updated or after all sub-windows of the window are updated. The second is called \emph{latest-window-serializable}, which only allows a read on the latest version of the window, i.e., after the window has been completely updated. Instead of imposing an event ordering, FlowDB\cite{Affetti:2017:FIS:3093742.3093929} enables developers to optionally ensure that the effects of transactions are the same as if they were executed sequentially (i.e., in the same order in which they started).}

\subcompact
\paragraph{Operation Ordering:}
\change{In many TSP systems, applications can be represented using dataflow models, such as directed acyclic graphs (DAGs), where operators are connected by data streams~\cite{tstream}. However, it is essential to acknowledge that in the TSP landscape, there may be other models that represent and process data streams using a different approach.\margi{R1Q41}Operation ordering refers to the sequence in which operators are executed in a TSP system, which impacts the correctness and efficiency of the system. Via operation ordering, 
the data that flows through the  pipeline will be processed correctly:
thereby contributing to the consistency and correctness prevalent in a TSP system. 
When an application is represented using a DAG, the operation ordering is determined by the directed edges between the operators.
Although the operation ordering may be expressed differently for alternative representations, 
the ordering property guarantees  the outcome will be the same.
Consider a scenario where a TSP system adopts a dataflow model for stream processing - the ordering property is inherently upheld in such a setup. Nevertheless, if a database system incorporates TSP, an added prerequisite for operation ordering becomes necessary to guarantee consistency and accuracy, as detailed in the work of S-Store~\cite{S-Store}.
}

\subcompact
\subsubsection{ACID Properties}
\change{TSP can offer traditional ACID guarantees~\cite{Bernstein:1996:PTP:261193} similar to those in relational databases.\margi{R1Q31}However, depending on the requirements of the specific application, TSP systems may need to relax or adapt these guarantees.} First, although \acid properties are strictly required in relational databases, they may be relaxed in stream processing scenarios~\cite{Affetti:2017:FIS:3093742.3093929,botan2012transactional,tstream,ACEP}.
Second, some streaming applications, do not require the \acid properties~\cite{ACEP}. 

\subcompact
\paragraph{Atomicity:} 
\change{Atomicity ensures that  transactions are either fully completed or aborted. In TSP systems, all operations within a transaction are either successfully processed together or not processed at all: thereby preventing partial updates that could lead to an inconsistent state. Atomicity in TSP varies depending on the transaction model. Traditional commit protocols, such as two-phase commit (2PC), ensure atomicity\margi{R1Q44}by coordinating commit or abort decisions among distributed participants. In contrast, sagas~\cite{sagas} allow the exposure of intermediate (uncommitted) state and require developers to define compensating actions for each operation, thereby offering a more flexible way to handle atomicity at the cost of strong isolation guarantees. Some TSP systems, like the one proposed by Wang et al.~\cite{ACEP}, relax atomicity in certain contexts, which enables developers to choose the desired consistency level. In such cases, alternative atomicity models, like sagas, can be adopted to balance the trade-offs between consistency, performance, and availability. Understanding these differences is crucial for designing TSP systems with appropriate atomicity guarantees.}

\subcompact
\paragraph{Consistency:}
Consistency ensures that a transaction moves the system from one consistent state to another consistent state~\cite{Vossen2016}. In the context of TSP systems, consistency guarantees that data is processed and updated according to the specified consistency model. TSP systems can provide different levels of consistency depending on the application requirements, such as eventual consistency~\cite{ACEP}, strong consistency~\cite{Affetti:2017:FIS:3093742.3093929}, or causal consistency~\cite{tstream}. It is essential to understand the trade-offs between various consistency levels, as stronger consistency guarantees may lead to increased complexity and reduced performance.

\subcompact
\paragraph{Isolation:}
Isolation prevents concurrent transactions from interfering with one another~\cite{Vossen2016}. In TSP systems, isolation is essential for ensuring that the output data remains consistent despite the concurrent execution of transactions. Different isolation levels can be provided by TSP systems, such as serializability, snapshot isolation, or read committed~\cite{Affetti:2017:FIS:3093742.3093929}. The choice of isolation level depends on the application requirements and the trade-off between consistency and performance. Some TSP systems may offer configurable isolation levels, allowing developers to adjust the isolation guarantees according to the application's specific needs.

\subcompact
\change{
\paragraph{Durability:}
Durability\margi{R1Q36}guarantees that once a transaction is committed, its changes are permanently stored in the system~\cite{Vossen2016}. In TSP systems, durability can be ensured through techniques, such as replication, logging, or checkpointing~\cite{Affetti:2017:FIS:3093742.3093929}. 
Durability plays a pivotal role in TSP systems, especially when recovery from failures and fault tolerance are considered. A system's durability influences the preservation of its state and output and governs the perceived behaviour from an external user's standpoint during a failure event. The recovery mechanism of specific TSP systems might entail replaying input streams to rebuild the state, which, due to variability in order and isolation presumptions, may not always restore an identical state. The extent of state preservation post-recovery can differ. To ensure durability, a TSP system typically needs to satisfy various properties, including input preservation, state maintenance, and output persistence. This is often achieved through strategies like replication, logging, or checkpointing, which assure fault tolerance and recoverability post-failure.
}

\subcompact
\subsubsection{State Management Properties}
Effective state management is essential in TSP systems to ensure data consistency, support the stateful processing of operations, and failure recovery. Among the primary state management properties are state types, access scope, state recovery, and the management of large states.

\subcompact
\paragraph{State Types:}
States in TSP systems can be categorized into two types: a) read-only state: This type of state involves applications that solely read data to obtain the information needed to process input events. Managing read-only state is relatively straightforward, as it does not involve updates or concurrency control concerns. b) read-write state: This type of state is updated and maintained as stream events are processed. Managing read-write state becomes particularly challenging when processing different input events that rely on reads or writes to the same state. Concurrency control issues can arise when more than one entity (e.g., threads) shares a state (e.g., intra-operator) and may concurrently modify it during execution.

\subcompact
\paragraph{Access Scope:}
Access scope defines the visibility and accessibility of state information within a TSP system. It dictates whether the state is shared globally, partitioned across parallel processing instances, or limited to a specific operator or group of operators. In the case of stream operators, they may need to maintain multiple states. For example, shared states can be the index structure of an input stream or some other user-defined data structure shared among \textit{threads} of the same operator, \textit{operators}, and \textit{queries}. Moreover, access scope impacts performance, consistency, and fault tolerance. 

\subcompact
\paragraph{State Recovery:}
State recovery refers to the ability of a TSP system to restore its state after a failure, thereby, ensuring that processing can resume without data loss or inconsistencies. This property is essential for providing fault tolerance and maintaining the overall reliability of a TSP system. State recovery can be achieved using techniques, such as checkpointing, logging, or state replication. Effective state recovery mechanisms must minimize the time required to restore  state and resume processing while ensuring that the recovered state is consistent with the pre-failure state.

\paragraph{Management of Large States:}
As the volume of data and the number of concurrent transactions grows, TSP systems must be able to manage large states effectively to maintain low-latency access and fault tolerance. To distribute shared mutable state across multiple nodes, techniques, such as data sharding, partitioning, and replication can be employed, to provide scalability. State partitioning involves distributing state information across multiple storage locations, improving scalability and parallelism. State compaction reduces the storage footprint by merging redundant or outdated state information. State checkpointing periodically saves snapshots of the state, enabling recovery after failures and ensuring durability. Additionally, TSP systems can use state eviction and expiration strategies, such as Least Recently Used (LRU) or Time-to-Live (TTL), to manage memory and storage resources efficiently. This ensures that the system can continue processing new transactions without running out of resources.

\subcompact
\subsubsection{Reliability and Delivery Guarantees}
Fault tolerance, durability, and deliverability are essential properties of TSP systems. These properties ensure that TSP systems can accurately process data and maintain their state despite potential issues, such as failures, duplicate messages, and out-of-order events. Fault tolerance and durability are essential to achieve exactly-once delivery semantics. Fault tolerance ensures a system can continue operating despite failures, while durability guarantees that committed transactions are stored permanently. Together, these properties enable the system to provide exactly-once semantics, thereby, ensuring each event is processed correctly and without duplication or omission, even when failures occur.

\subcompact
\paragraph{Delivery Guarantees:}
Delivery guarantees in TSP systems dictate how input events are processed, when failures, duplicate messages, or out-of-order events occur. Common delivery guarantees in stream processing systems include at-most-once, at-least-once, and exactly-once processing. Exactly-once processing is often the most desirable for TSP systems, ensuring each event is processed exactly once, regardless of failures or other issues during processing. Thus, even in the presence of failures, the system can ensure no duplicates or missing events occur. 

\subcompact
\paragraph{Fault Tolerance:}
Since streaming applications typically run indefinitely, the probability of service disruptions due to unexpected system failures increases. A fault-tolerant TSP system persists when processing events and adheres to delivery guarantees, even in the face of failures. In case of a failure, a fault-tolerant system must recover and resume processing without violating delivery guarantees. To enable rapid recovery from failures and minimize data loss, fault tolerance approaches, such as state replication, checkpointing, and log-based recovery are employed. 

\subcompact
\paragraph{Durability:}
Durability involves persistent data storage, to ensure its availability for retrieval. This property is crucial for upholding delivery guarantees, as it enables the system to maintain desired processing semantics. For instance, in an exactly-once processing scenario, the system must ensure that each input event is processed and committed exactly once, necessitating durable storage for both the input events and corresponding state changes.

\subcompact
\paragraph{CAP Theorem:}
The CAP theorem, foundational in distributed systems, states a system can't simultaneously provide consistency, availability, and partition tolerance. This is of importance to Transactional Stream Processing (TSP) systems, which often must strike a balance among these properties due to distributed data processing and storage. Consistency and durability both aim to maintain a uniform view of the system's state. For TSP systems, these are crucial for preserving data integrity and providing reliable transactional guarantees. Availability, ensuring every system request receives a response, is vital for continuous, real-time processing in TSP systems. Partition tolerance, akin to fault tolerance, allows a system to function even during communication breakdowns between nodes, a key requirement in TSP systems for ensuring uninterrupted processing. Given the CAP theorem's constraints, TSP systems may need to prioritize among these properties based on specific use cases. For instance, a TSP system prioritizing consistency and partition tolerance might compromise availability, leading to longer response times. Conversely, one focusing on availability and partition tolerance may occasionally return stale data but provides fast responses and continuous operation.
\color{black}

\begin{remark}[Beyond Exactly-Once Guarantee]
TSP systems may face unique challenges requiring more stringent delivery guarantees than the exactly-once guarantee typically found in many SPEs. Specifically, TSP systems must replay failed tuples in the exact timestamp sequence of their triggering input events and prevent duplicate message processing. This is crucial because results depend on the local state of an operator and the time ordering of input streams.

One approach to achieving this advanced level of delivery guarantee involves checkpointing or archiving each input event before processing and sequentially replaying them in case of failure~\cite{Chandrasekaran2004RemembranceOS}. While this method provides the desired guarantee, it incurs significant overhead, making it unsuitable for many TSP systems. Consequently, further research is required to identify more efficient mechanisms that can satisfy the delivery guarantee needs of TSP systems while minimizing performance overhead.    
\end{remark}

\subcompact
\subsection{Design Aspects of TSP}
\label{sec:design}
The design aspects of TSP systems focus on the critical components required to create a functional and efficient TSP system. These components include the implementation of transactions, determining transaction boundaries, executing transactions, delivering guarantees, and managing the state. Understanding these design aspects is crucial for evaluating the suitability of different TSP systems for specific application requirements and providing insights into various design choices and trade-offs.

\subcompact
\subsubsection{Implementing Transactions}
To actualize a transaction model over data streams, three primary approaches have been proposed:\margi{R1Q26, R3Q6}embedding stream processing operations into transactions (i.e., \emph{unified transactions}), embedding transaction processing into stream processing (i.e., \emph{embedded transactions}), or combining transaction processing and stream processing (i.e., \emph{state transactions}). Let us discuss each of these in turn.

\paragraph{Approach 1: Unified Transactions.}
Unified transactions integrate stream processing and transaction processing by transforming data streams into time-varying relations~\cite{motwani2002query,stream2003stream,Arasu:2006:CCQ:1146461.1146463}. By treating data streams and relational data uniformly, TSP systems can leverage existing relational data management techniques for transactional consistency and correctness. This approach enables us to define a transaction that can handle both stream processing and transactional aspects within a single framework, thereby, allowing flexibility in defining the transaction scope and granularity.

\begin{definition}[Unified Transaction]
Let $D = {d_1, d_2, ..., d_n}$ be a dataset containing both data streams and relational data, and $O = {o_1, o_2, ..., o_m}$ be a set of processing operations. A unified transaction $T_i$ is a tuple $(D_i, O_i)$, where $D_i \subseteq D$ and $O_i \subseteq O$. In this approach, both stream processing and transactional aspects are handled within a single framework, allowing flexibility in defining the scope and granularity of transactions. 
\label{def:unified_transaction}
\end{definition}

Several studies have explored the implementation of continuous queries as sequences of one-time queries, which are executed as a result of data source modifications or periodic execution~\cite{botan2012transactional, S-Store, oyamada2011efficient, Oyamada13}. These studies illustrate the various methods employed to unify stream processing and transaction processing, so as to enable the use of a single execution engine for both types of operations.
The design space for unified transactions consists of several aspects, including data modelling, query execution, and transaction management.

\textbf{Data Modeling.} 
In order to transform data streams into relational data, data streams are represented as time-varying relations.
To indicate the time-varying nature of data, both input and output streams are treated as continuous relations with a timestamp attribute. This uniform representation enables the use of relational algebra and SQL-like queries to process data streams and simplifies the integration of stream processing with existing relational database systems.

\textbf{Query Execution.} Queries over data streams are often continuous: they are evaluated over an unbounded sequence of input data. Unified transactions represent continuous queries as a sequence of one-time queries triggered by data source modifications or periodic execution. This approach allows for the reuse of existing query processing techniques from relational databases while ensuring the continuous nature of stream processing is maintained.

\textbf{Transaction Management.} 
To ensure transactional consistency and correctness, the unified transactions approach relies on relational data management techniques. In particular, concurrency control and recovery mechanisms, such as two-phase locking and logging, to provide isolation, atomicity, and durability guarantees.

The unified transactions approach simplifies system architecture and leverages well-established relational data management techniques. However, this approach may introduce additional overhead due to the transformation of data streams into relational data and may not be suitable for all TSP systems, particularly those with strict latency requirements or complex stream processing needs. Developers must carefully evaluate the trade-offs and requirements of their specific application to determine whether this approach is appropriate for their TSP system.

\paragraph{Approach 2: Embedded Transactions.}
The embedded transaction approach integrates transaction processing directly into the stream processing pipeline. Instead of transforming data streams into relational data, transactions are managed and processed within the context of stream processing. This approach involves event-driven processing, state management, and transactional guarantees.

\begin{definition}[Embedded Transaction]
Let $S = {s_1, s_2, ..., s_n}$ be a data stream, and $O = {o_1, o_2, ..., o_m}$ be a set of stream processing operations. An embedded transaction $T_i$ is a tuple $(s_i, O_i)$, where $s_i \in S$ and $O_i \subseteq O$. Here, each incoming data item $s_i$ is treated as a transaction, allowing real-time processing while maintaining consistency and reliability guarantees.
\label{def:embedded_transaction}
\end{definition}
\margi{R1Q19, R3Q9}

Several studies, including those exploring incremental continuous query processing with isolation guarantees~\cite{Shaikh16} have explored the embedded transactions approach. For instance, Shaikh et al~\cite{Shaikh16} propose a model that treats each incoming data item as a transaction. This approach allows for real-time processing of data streams while maintaining consistency and reliability guarantees. However, potential issues include inconsistencies in the output due to updates during join processing. To address this, they propose a mechanism for maintaining consistency in the presence of concurrent updates, using snapshot isolation and optimistic concurrency control.

Key aspects of embedded transactions involve processing events as they arrive, managing state information between processing steps, and incorporating data partitioning, replication, and checkpointing to achieve transactional guarantees. Next, we examine each of these aspects in turn.

\textbf{Event-Driven Processing.}
The embedded transaction approach employs event-driven processing, where each event in the data stream triggers one or more stream processing operations. This allows for low-latency processing and maintains the temporal order of data streams.

\textbf{State Management.} 
State management is crucial in the embedded approach since it enables the retention and manipulation of information between processing steps. State management varies across systems, with some using distributed storage systems or in-memory data structures for efficient state management.

\textbf{Transactional Guarantees.} 
The embedded transaction approach provides transactional guarantees, such as consistency, isolation, and durability within the stream processing pipeline. To achieve these guarantees, they employ techniques, such as data partitioning, replication, and checkpointing.

The embedded transaction approach offers benefits, such as low-latency processing and the native handling of complex stream processing tasks. Nevertheless, implementing transactional guarantees within the stream processing pipeline could necessitate substantial efforts. Furthermore, this approach might not be optimal for applications necessitating integration with established relational databases or traditional transaction processing systems. Developers should carefully assess the requirements and constraints of their specific application to determine whether this approach is appropriate for their TSP system.

\paragraph{Approach 3: State Transactions.}
The state transaction approach merges transaction processing and stream processing within a single system and handles state access operations as transactions. This enables TSP systems to provide transactional guarantees while processing unbounded data streams and ensure correctness via transactional semantics and the modelling of state accesses as \emph{state transactions}.

\begin{definition}[State Transaction]
Let $S = {s_1, s_2, ..., s_n}$ be a data stream, $R = {r_1, r_2, ..., r_p}$ be a set of shared mutable states, and $O = {o_1, o_2, ..., o_m}$ be a set of processing operations. A state transaction $T_i$ is a tuple $(S_i, R_i, O_i)$, where $S_i \subseteq S$, $R_i \subseteq R$, and $O_i \subseteq O$. All operations within the same transaction have the same timestamp, denoted as $ts$.
\label{def:state_transaction}
\end{definition}

State transactions focus on managing shared mutable states through transactions. Key aspects include managing state information between processing steps and decoupling transaction processing from stream processing. Design considerations include dataflow models, state access operations, coordination mechanisms, and fault tolerance and recovery. Next, we examine each of these design considerations in turn.

\textbf{Dataflow Models.} 
The state transaction approach typically uses dataflow models to process data streams consisting of interconnected stateful operators. These operators process events, update state, and produce output events, which is amenable for parallel and distributed processing.

\textbf{State Access Operations.} 
State transactions treat state access operations as transactions, where each operation is associated with a unique timestamp. This enables complex processing tasks and transactional guarantees, like consistency, isolation, and durability.

\textbf{Coordination Mechanisms.} 
The state transaction approach coordinates state transactions across the dataflow pipeline using mechanisms, such as two-phase commit protocols, timestamp-based ordering, or conflict resolution strategies, to maintain consistency and isolation.

\textbf{Fault Tolerance and Recovery.} 
The state transaction approach provides fault tolerance and recovery mechanisms, such as replication, checkpointing, and logging, to ensure durability and resilience against failures.

The state transaction approach integrates transaction processing into the stream processing pipeline, handles complex processing tasks, and provides transactional guarantees. However, this approach may introduce additional complexity due to the use of coordination and fault tolerance mechanisms. Developers should carefully assess their application's requirements and weigh the benefits against the potential complexities introduced by this approach.

\subcompact
\subsubsection{Determining Transaction Boundaries}
Determining transaction boundaries is an essential aspect of TSP system design, as it defines which operations are grouped into transactions for atomicity.
It turns out that establishing transaction boundaries over streams can be quite flexible, particularly for state transaction implementation. First, various conditions can initiate a transaction (i.e., \emph{triggering unit}), such as per input event or per batch of events with a common timestamp. Second, different entities can generate a transaction (i.e., \emph{generating unit}), such as per operator and per query. Additionally, transactions can spawn other transactions (i.e., \emph{transaction spawning}). Let us examine each of these settings in turn.

\paragraph{Setting 1: Triggering Units.}
TSP systems diverge from traditional databases in that they rely on incoming streaming events rather than user requests to initiate transactions. The granularity of transaction boundaries is defined by various types of triggering units, including time-based, batch event-based, single event-based, and user-defined triggers. Next, we describe each of these in turn.

\textbf{Time-based Triggers.} 
Time-based triggers\margi{R1Q24, R1Q25}refer to transactional models in which the transaction boundaries are determined by time intervals. These intervals can be either fixed or dynamically adjusted based on the application requirements or the characteristics of the data streams. It often assumes that events with a common timestamp are executed atomically to ensure progress correctness. This approach is employed in both academic projects, such as STREAM~\cite{stream2003stream,Arasu:2006:CCQ:1146461.1146463} and commercial products, like Coral8~\cite{Coral8}. Time-based triggers are suitable for the  concurrent aggregation of sliding windows, the association of transaction boundaries with window boundaries, and the management of long-running queries with specified re-execution frequencies~\cite{golab2,DSPCC,conway2008cisc}.

\textbf{Batch Event-based Triggers.} 
These transactions are triggered when a batch of events with a shared characteristic (e.g., common timestamp, originating from the same stream) are processed. Batch event-based triggers are used in DataCell~\cite{Liarou2009DataCellBA,liarou2009exploiting}, S-Store~\cite{S-Store}, and Chen et al.'s cycle-based transaction model~\cite{Chen10,chen2011query}.They can handle large volumes of data and provide consistent processing across multiple streams.

\textbf{Single Event-based Triggers.} In this type, a transaction is triggered for each incoming event. Single event-based triggers ensure fine-grained control and consistency on an event-by-event basis. They have been implemented in various systems, such as Aurora and Borealis~\cite{Aurora,abadi2005design}, ACEP~\cite{ACEP}, SPASS~\cite{SPASS}, and TStream~\cite{tstream}. This type of transaction is suitable for applications requiring strict consistency guarantees and low-latency processing.

\textbf{User-defined Triggers.} 
In this type, users can define custom triggering conditions based on their specific application requirements, which offers flexibility when establishing transaction boundaries and declaring processing guarantees. Botan et al.\cite{botan2012transactional} and Chen et al.\cite{chen2018streamdb} demonstrate the use of user-defined transactions in their respective systems.


\paragraph{Setting 2: Generating Units.}
While triggering units determine ``when'' a transaction is created, generating units focus on ``who'' generates a transaction. Transactions can be generated by user clients directly or through continuous queries on a per-query or per-operator basis. The design space for generating units includes query-based, operator-based, and user-defined generators. Next, we describe each unit type in turn.

\textbf{Query-based Generator.} 
These transactions group operations involved in the one-time execution of an entire query \cite{stream2003stream,Arasu:2006:CCQ:1146461.1146463}. Early Stream Processing Engines (SPEs) employed query-based triggers for the interactive processing of both relational and streaming data, such as the STREAM project \cite{stream2003stream,Arasu:2006:CCQ:1146461.1146463}. This type simplifies the transactional model, making it easier to manage and understand. However, it may lack flexibility in some cases, where individual operators within a query need to have separate transactional boundaries or different isolation levels.

\textbf{Operator-based Triggers.} 
In this type, each operator in a query generates its own transactions \cite{S-Store,tstream}. Examples of this design include S-Store \cite{S-Store} and TStream \cite{tstream}. Operator-based triggers provide a finer level of granularity and flexibility compared to query-based triggers, which enables more precise control over shared states in streaming dataflow graphs. This can lead to better performance and resource utilization in certain scenarios. However, this increased flexibility may also result in potential conflicts or dilemmas, such as deadlocks and contention, which may require additional mechanisms to resolve \cite{S-Store,tstream}.

\textbf{User-defined Triggers.} 
Some applications may require ad-hoc transactional queries or user-driven transactions during stream processing \cite{AFFETTI202065,Affetti:2017:FIS:3093742.3093929,chen2018streamdb}. User-defined transactions allow users to specify where consistency needs to be enforced and which consistency constraints are required, as demonstrated in the work of Affetti et al. \cite{AFFETTI202065,Affetti:2017:FIS:3093742.3093929}. This type grants more control to the user, thereby, allowing them to tailor transactional semantics to their specific needs. However, this flexibility can make system-level optimizations more challenging, as the transaction types are not known in advance. Additionally, users bear the responsibility of ensuring that the system is free of any dilemmas or conflicts \cite{AFFETTI202065,Affetti:2017:FIS:3093742.3093929}.


\paragraph{Setting 3: Transaction Spawning.}
Transactions can also spawn, i.e., trigger and generate other transactions. 
This is particularly useful in service-oriented architectures, where the basic premise is to treat all functionalities as services and compose and execute them, according to the user or application-specific requirements~\cite{DEXA16}. Treating each service execution as a transaction and requiring atomic execution of those transactions have been found to be very helpful in this process. The execution of service compositions yields composite transactions~\cite{TM16}, where a transaction is an execution of a service~\cite{TIOT}.
Subsequently, a transaction spawning consists of a nonempty set of services. Some of these have executions that are continuous and others may spawn new transactions. 

\subcompact
\subsubsection{Executing Transactions}
Executing transactions involves processing the operations within a transaction according to the defined transaction boundaries and ensuring that the system maintains the required properties including the \acid properties and also the streaming properties.
\change{
Table~\ref{tab:execution} summarizes\margi{R1Q49, R3Q14}the execution mechanisms adopted by relevant systems. These can be classified into five approach types: \emph{single-version lock-based}, \emph{multi-version lock-based}, \emph{static partition-based}, \emph{dynamic partition-based}, and \emph{optimistic}. Note that, ``C'' in ACID refers to the preservation of integrity constraints, independent of how transactions are integrated with stream processing.
}

\begin{table}[t]
\centering
\caption{Execution mechanisms of \tsp systems.}
\margi{R1Q49}
\label{tab:execution}
\resizebox{0.5\textwidth}{!}{%
\begin{tabular}{|L{2.7cm}|L{1.8cm}|L{4.8cm}|}
\hline
\textbf{Works} & \textbf{Approach}                             & \textbf{Key Notes} \\ \hline
STREAM~\cite{stream2003stream}, Wang et al.~\cite{ACEP}, Oyamada et al.~\cite{oyamada2011efficient}, FlowDB/TSpoon~\cite{Affetti:2017:FIS:3093742.3093929, AFFETTI202065} & single-version lock-based    & Each state is maintained with a single copy; concurrent access is regulated by exclusive locks with an event ordering guarantee        \\ \hline
Wang et al.~\cite{ACEP}       & multi-version lock-based     & Each state is maintained with multiple copies; concurrent access is regulated by read-write locks with an event ordering guarantee          \\ \hline
S-Store~\cite{S-Store} & state partition-based        & Pre-partition states into disjoint partitions, and regulate concurrent access to each partition, similar to the single-version lock-based approach       \\ \hline
TStream~\cite{tstream}, MorphStream~\cite{mao2023morphstream} & transaction partition-based  &   Dynamically partition and regroup state transactions to avoid conflicts\\ \hline
Golab et al.~\cite{golab2}, FlowDB/TSpoon~\cite{AFFETTI202065,Affetti:2017:FIS:3093742.3093929}       & optimistic                   & Optimistically schedule concurrent state transactions and abort transactions if conflicts arise   \\ \hline
Several prior works~\cite{abadi2005design,botan2012transactional,Chen11,Chen10,Gtze2019SnapshotIF} & snapshot isolation            &       Implement the semantics of database transactions to defined segments of data streams, thereby assuring snapshot isolation during the processing of these segments  \\ \hline
\end{tabular}%
}
\end{table}

\paragraph{Lock-based Approaches:}
Lock-based approaches ensure the correct execution of transactions by controlling access to shared resources. Using locks to protect shared states, these methods prevent concurrent access and maintain consistency. Lock-based approaches can be classified into two main types:

\emph{a) Single-version lock-based:} 
These approaches utilize a single version of the shared state and apply locks to ensure proper transaction execution. The challenge lies in balancing synchronization and performance without causing excessive contention or delays in transaction processing. We discuss three notable examples of single-version lock-based approaches as follows.
 
 In STREAM~\cite{stream2003stream}, synopses enable different operators to share common states. To guarantee that operators view the correct version of a state, the system must track the progress of each stub and present the appropriate view (i.e., a subset of tuples) to each stub. This is achieved through a local timestamp-based execution model with a global schedule that coordinates the successive execution of individual operators via time slot assignments. Batches of tuples with the same timestamp are executed atomically to ensure progress correctness, with a simple lock-based transactional processing mechanism implicitly involved.

An earlier study by Wang et al.~\cite{ACEP} describes a strict two-phase locking (S2PL)-based algorithm that allows multiple state transactions to run concurrently while maintaining both ACID and streaming properties. Unlike the original S2PL~\cite{Bernstein:1996:PTP:261193} algorithm, Wang et al.~\cite{ACEP} lock each transaction ahead of all query and rule processing. In this process, each transaction's timestamp is compared against a monotonically increasing counter to ensure that the transaction with the smallest timestamp always obtains a lock first, thereby guaranteeing access to the proper state sequence. Once lock insertion is complete, the system increases the counter and allows the next transaction to proceed, regardless of whether the transaction was fully processed. To fulfil event ordering constraints, read or write locks are strictly invoked in their triggering event order. However, the locking mechanism must synchronize the execution for every single input event, which may negatively impact system performance.

Oyamada et al.~\cite{oyamada2011efficient} propose three pessimistic transaction execution algorithms: synchronous transaction sequence invocation (STSI), asynchronous transaction sequence invocation (ATSI), and order-preserving asynchronous transaction sequence invocation (OPATSI). STSI processes transactions triggered by event streams one at a time, with execution results naturally generated following the event arrival sequence. ATSI removes the blocking behaviour of STSI by asynchronously spawning new threads that wait for the transaction to complete. OPATSI extends ATSI through a priority queue to further guarantee the order of the results.
 
\emph{b) Multi-version lock-based:} 
These approaches employ multiple versions of shared states and use locks to control access to different state versions. The main challenge is ensuring that the correct state version is accessed while avoiding outdated writes. 

A notable example is Wang et al.~\cite{ACEP}, who propose an algorithm called LWM (Low-Water-Mark), which relies on the multi-versioning of shared states. LWM leverages a global synchronization primitive to guard the transaction processing sequence: write operations must be performed monotonically in event order, but read operations are allowed to execute as long as they can read the correct version of the data (i.e., its timestamp is earlier than the LWM). The key differences between LWM and the traditional multi-version concurrency control (MVCC) scheme are twofold. First, MVCC aborts and then restarts a transaction when an outdated write occurs, while LWM ensures that writes are permitted strictly in their timestamp sequence, preventing outdated writes. Second, MVCC assumes that the timestamp of a transaction is system-generated upon receipt, whereas LWM sets the timestamp of a transaction to the triggering event. This distinction enables LWM to maintain a more event-driven approach to transaction management, better aligning with the streaming nature of TSP systems.

In summary, lock-based approaches to transaction execution in TSP systems offer various methods for managing access to shared resources and maintaining consistency. While single-version lock-based approaches focus on balancing synchronization and performance within a single shared state, multi-version lock-based approaches provide greater flexibility by managing multiple versions of shared states. Both types of approaches present their own challenges and trade-offs, and the choice of approach depends on the specific requirements and characteristics of the TSP system being implemented.

\paragraph{Partition-based Approaches:}
Partition-based approaches to transaction execution in TSP systems involve dividing the internal states or transactions into smaller units, which can then be executed in parallel or with reduced contention. These methods aim to improve performance while maintaining consistency and adhering to event order constraints. There are two primary types of partition-based approaches:

\emph{a) Static partition-based:}
These approaches divide the internal states of streaming applications into disjoint partitions and use partition-level locks to synchronize access. This approach is suitable for transactions that can be perfectly partitioned into disjoint groups.

For example, S-Store~\cite{S-Store} splits the streaming application's internal states into multiple disjoint partitions. The computation on each sub-partition is performed by a single thread. To guarantee state consistency, S-Store uses partition-level locks to synchronize access.
However, the state partition-based approach only performs well on transactions that can be perfectly partitioned into disjoint groups, given that acquiring partition-level locks on cross-partition states significantly impacts performance due to the overhead.

\emph{b) Dynamic partition-based:} These approaches involve decomposing transactions into smaller steps and executing them in parallel to improve performance while ensuring serializability and meeting event order constraints (e.g., TStream~\cite{tstream} and MorphStream~\cite{mao2023morphstream}).

The \emph{sagas} model~\cite{sagas} allows a transaction to be split into several smaller steps, each of which executes as a transaction with an associated compensating transaction.
Either all steps are executed or in a partial execution compensating transactions are executed for steps that are completed.
Thus, isolation is relaxed in the original transaction and delegated to the individual steps.
It exposes an intermediate (uncommitted) state and requires developers to define compensating actions.
A similar idea of splitting transactions has been adopted in \tsp systems such as TStream~\cite{tstream} and MorphStream~\cite{mao2023morphstream} but does not expose uncommitted states and hence does not require compensating actions.

In particular, TStream~\cite{tstream} is a recently proposed \tsp system that adopts transaction decomposition to improve stream transaction processing performance on modern multicore processors. Despite the relaxed isolation properties, TStream ensures serializability, as all conflicting operations (being decomposed from the original transactions) are executed sequentially as determined by the event sequence. The successor of TStream~\cite{tstream}, MorphStream pushes the idea further and proposes cost-model guided dynamic transaction decomposition and scheduling to further improve the system performance.

\paragraph{Optimistic Approach:}
Optimistic approaches avoid locking resources by employing timestamps and conflict detection mechanisms to maintain transaction consistency at the desired isolation level, aborting and rescheduling transactions when necessary.
These approaches handle transactions by predicting the order of events or by performing speculative execution to improve system performance. The challenge is to ensure that speculation is accurate and efficiently manages rollback or recovery when needed.

Golab et al.\cite{golab2} present a scheduler targeting \emph{window serializable} properties, which optimistically executes window movements and utilizes serialization graph testing (SGT) to abort any \emph{read-only} transactions causing read-write conflicts. A conflict-serializable schedule is achieved if the precedence graph remains acyclic. They also suggest reordering read operations within transactions to minimize the number of aborted transactions, thereby improving the schedule. FlowDB/TSpoon\cite{AFFETTI202065, Affetti:2017:FIS:3093742.3093929} propose an optimistic timestamp-based protocol that refrains from locking resources and instead uses timestamps to ensure transactions consistently read or update versions aligned with the desired isolation level. If this is not feasible, transactions are aborted and rescheduled for execution. This approach aims to minimize contention and improve performance by avoiding lock-based mechanisms while still maintaining the necessary consistency and isolation requirements.
 
\paragraph{Snapshot Isolation Approach:}
These approaches employ snapshot isolation to split a stream into a sequence of bounded chunks and apply database transaction semantics to each chunk. Processing a sequence of data chunks generates a sequence of state snapshots. By storing multiple versions of values as commit and delete timestamps, readers can access the latest version of a state, ensuring consistency and isolation among concurrent transactions.

A number of TSP systems employ snapshot isolation~\cite{abadi2005design,botan2012transactional,Chen11,Chen10,Gtze2019SnapshotIF}, aiming to split a stream into a sequence of bounded chunks and apply the semantics of a database transaction to each chunk. By putting the operation on a data chunk within a transaction boundary, a state snapshot is produced. In this way, processing a sequence of data chunks generates a sequence of state snapshots.
For example, Götze and Sattler~\cite{Gtze2019SnapshotIF} present a snapshot isolation approach for TSP. Each state has multiple versions of values, each stored as a \emph{commit timestamp}, \emph{delete timestamp}, and \emph{value}. Consequently, readers can access the latest version of a state using the commit and delete timestamps. This approach provides consistency and isolation among concurrent transactions while avoiding the need for locking mechanisms, which can improve system performance.

\subcompact
\subsubsection{Ensuring Delivery Guarantees}
In this subsection, we explore various design aspects of TSP systems that help ensure reliability and delivery guarantees. We discuss strategies for achieving ACID properties and streaming properties under failures and their implications on TSP system design. For a comprehensive survey on fault tolerance mechanisms in SPEs, refer to~\cite{state_survey}. While modern SPEs usually offer fault-tolerance mechanisms while ensuring various delivery guarantees, they may not always fulfil the requirements of \tsp due to the combined need to satisfy \acid and streaming properties.

\paragraph{Achieving \acid Properties:}
\change{In the event of a failure, \tsp systems generally need to recover all states, 
including input/output streams, operator states, and shared mutable states. This ensures committed transactions remain stable, while uncommitted transactions do not impact this state. Transactions that have started, but have not yet been committed should be undone upon failure and reinvoked with the correct input parameters once the system is stable again. This necessitates an upstream backup and an undo/redo mechanism akin to an \acid-compliant database.
}

For instance, \tsp systems must guarantee \emph{atomicity} when updating shared states, even under failures. An atomic transaction ensures a commit either fully completes the entire operation or, in cases of failure (e.g., system failures or transaction aborts), rolls back the database (or shared states in \tsp) to its pre-commit state. Journaling or logging in database systems mainly accomplish atomicity, while distributed database systems require additional atomic commit protocols to ensure atomicity.
Regrettably, most prior works on \tsp either do not explicitly mention their mechanisms to ensure atomicity under failure~\cite{tstream,Affetti:2017:FIS:3093742.3093929} or rely on mechanisms provided by their storage systems (e.g., traditional database systems~\cite{S-Store}). Making this more transparent could help users better understand which properties are not guaranteed when employing a \tsp system in practice.

\paragraph{Achieving Streaming Properties:}
To satisfy streaming properties further, the recovered states in \tsp systems should be equivalent to the one under construction when no failure occurred. Achieving this requires an order-aware recovery mechanism~\cite{Clonos}. However, the commonly adopted recovery operation in modern SPEs, particularly the parallel recovery operation, might result in different transactional states due to the absence of guarantees on the event processing sequence during recovery. To the best of our knowledge, there is still no in-depth study on designing efficient fault tolerance mechanisms for \tsp systems.

\change{
Additionally,\margi{R1Q34}distributed SPEs often adopt a form of eventual consistency~\cite{10.1145/2460276.2462076} to ensure high availability, as states exposed to the external world are expressed as output streams, and instant consistency of the global system state is hidden from users. Eventual consistency informally guarantees that if no new updates are made to a given data item, all accesses to that item will eventually return the last updated value. However, it is unclear how such a relaxed model (i.e., eventual consistency) can be applied to \tsp to achieve fault tolerance and high availability, given that shared mutable states (or their snapshot) need to be immediately visible and queryable both internally and externally to the system.
}

\subcompact
\subsubsection{Implementing State Management}
\change{State\margi{R1Q56}management is a crucial aspect of transactional stream processing (TSP) systems, as it enables the coordination of concurrent transactions, maintains consistency, and provides fault tolerance~\cite{tstream, millwheel}. The design space for state management in TSP systems can be characterized by several dimensions, such as access scope, storage model, data manipulation statements, and state management strategies. Next, we delve into these dimensions and explore their implications for system design and optimization.}

\paragraph{Access Scope:}
The access scope of state management ranges from intra-operator to inter-systems. Depending on the application's requirements, TSP systems may need to manage state locally within a processing node or share state across multiple nodes or even external systems \cite{Gtze2019SnapshotIF, flink}. It is worth noting that when OLTP workloads are implemented in a TSP system, the access scope of a shared state is within a transaction, which can be attributed to a single operator or multiple operators.
\emph{a) Intra-operator state management} focuses on maintaining state among instances of a single operator, making it suitable for applications with localized data access patterns and minimal coordination requirements~\cite{golab2}. 
\emph{b) Inter-operator state management} involves sharing state across multiple operators within the same query/system~\cite{tstream,Affetti:2017:FIS:3093742.3093929}. This approach is particularly relevant for applications that require coordination among different operators.
\emph{c) Inter-system/global state management} extends the scope of state sharing even further, enabling TSP systems to exchange state information with external systems, such as other stream processors, databases, or distributed file systems~\cite{meehan2017data}. This approach allows TSP systems to leverage the capabilities of external systems, such as query processing or storage, and can facilitate seamless integration with existing data processing pipelines. However, managing state across system boundaries can introduce additional complexity, latency, and potential consistency issues.

\paragraph{Storage Models:}
There are two primary storage models for implementing state management in TSP systems: relations and key-value pairs. Each has its trade-offs and implications for system design and optimization.

\emph{a) Relations:} In this model, states are represented as time-varying relations that map a time domain to a finite but unbounded bag of tuples adhering to a relational schema \cite{stream2003stream, S-Store}. This approach leverages well-developed techniques from relational databases, such as persistence and recovery mechanisms. Storing states as relations can help minimize system complexity, especially when a foreign key constraint is required in TSP~\cite{meehan2017data}. However, incorporating time into the relational model can add complexity to query processing and optimization \cite{stream2003stream}.

Representative examples include STREAM, S-Store, and TStream.
STREAM~\cite{stream2003stream} represents the state as a time-varying relation, mapping a time domain to a finite, but unbounded bag of tuples adhering to the relational schema. To treat relational and streaming data uniformly, STREAM introduces two operations: \emph{To\_Table} to convert streaming data to relational data, and \emph{To\_Stream} to convert relational data to streaming data.
S-Store~\cite{S-Store} does not implement its own state management component, but instead relies on H-Store~\cite{HStore} to ensure the transactional properties of shared states represented as relations. TStream~\cite{tstream} uses the Cavalia relational database~\cite{Wu:2017:EEI:3067421.3067427} to support the storage of shared states.

\emph{b) Key-Value Pairs:}
In this model, states are represented as key-value pairs, which simplifies the design of TSP systems \cite{millwheel, flink}. This approach is suitable for scenarios that mainly require select and update statements for manipulating shared states during stream processing \cite{AFFETTI202065, mao2023morphstream}. However, it may not be the best choice for applications that require more complex data manipulation operations or constraints, such as foreign key constraints \cite{meehan2017data}.

Representative examples include MillWheel~\cite{millwheel}, Flink with RocksDB, AIM (Analytics in Motion)\cite{AIM15}, FlowDBMS/TSpoon\cite{AFFETTI202065,Affetti:2017:FIS:3093742.3093929}, and TStream/MorphStream~\cite{tstream,mao2023morphstream}.
MillWheel~\cite{millwheel} maintains state as an opaque byte string on a per-key basis, with users implementing serialization and deserialization methods. The persistent state is backed by a replicated and highly available data store, such as Bigtable~\cite{chang2008bigtable} or Spanner~\cite{corbett2013spanner}, ensuring data integrity and transparency for the end user.
Flink~\cite{flink} relies on an LSM-based key-value store engine called RocksDB~\cite{rocksdb} to support shared queryable state. G{"o}tze and Sattler~\cite{Gtze2019SnapshotIF} also adopt a key-value store for transactional state representation, using multi-version concurrency control, where each state (i.e., key) has multiple \emph{commit timestamps}, \emph{delete timestamps}, or \emph{values}.

AIM~\cite{AIM15} represents state in a distributed in-memory key-value store, where nodes store system state as horizontally-partitioned data in a \emph{ColumnMap} layout. The \emph{Analytics Matrix} system state provides a materialized view of numerous aggregates for each individual customer (subscriber). When an event arrives in an SPE, the corresponding record in the \emph{Analytics Matrix} is updated atomically.
In FlowDBMS/TSpoon~\cite{AFFETTI202065,Affetti:2017:FIS:3093742.3093929}, a key-value store is employed, with the state maintained by a special type of stateful stream operator called the \emph{state operator}. All state access requests must be routed to and subsequently handled by state operators defined in the application.

\paragraph{Data Manipulation Statements:}
TSP systems need to define and support different data manipulation statements employed in applications that constrain both system design and potential optimizations. These statements may include operations such as insert, update, delete, and query, which must be executed efficiently and consistently in the context of transactional stream processing.

Storing shared states as relations could be a reasonable choice of system design when applications require \emph{insert} (I) or \emph{delete} (D) statements and need to maintain foreign key constraints, such as in the case of streaming ETL~\cite{meehan2017data}. However, when applications only need \emph{select} (S) and \emph{update} (U) statements for manipulating shared states during stream processing, storing shared states as vanilla key-value pairs is sufficient and simplifies the design of \tsp systems. Specific optimizations should be adopted by the \tsp systems according to application needs.

\paragraph{State Management Strategies:}
The choice of state management strategy can significantly impact system performance, fault tolerance, and scalability \cite{millwheel, flink, HStore}. There are three main strategies for managing state in TSP systems:
\emph{a) In-memory state:} This strategy maintains state within the processing nodes' memory, enabling low-latency access. However, it can be limited by available memory and may require replication and distributed transactions for fault tolerance and consistency guarantees \cite{AIM15,tstream}.
\emph{b) External state stores:} In this strategy, state is stored in external systems, such as transactional databases or distributed key-value stores with transactional support \cite{Gtze2019SnapshotIF, rocksdb}. This approach allows for improved fault tolerance, consistency guarantees, and scalability but may introduce additional latency \cite{chang2008bigtable, corbett2013spanner}.
\emph{c) Hybrid state management:} This approach combines the advantages of in-memory state and external state stores, using in-memory caching to minimize latency and external transactional storage for fault tolerance, consistency guarantees, and scalability \cite{AFFETTI202065, Affetti:2017:FIS:3093742.3093929}.


\begin{remark}[Failure of Concurrency Control Protocols]
    Conventional concurrency control (CC) protocols, which are widely used in OLTP database systems, fail to guarantee the properties of TSP Systems. To illustrate why, we use a conventional timestamp-ordering concurrency control (T/O CC)~\cite{Bernstein:1981:CCD:356842.356846} as an example. Similar discussions can be also found in a prior
    work~\cite{ACEP}.
    Let $txn_1=write(k1,v1)$ and $txn_2=read(k1)$ be two distinct transactions. 
    For simplicity, let us assume that there are only these two transactions in the system. If $txn_2.ts > txn_1.ts$, then both $txn$ will be successfully committed. 
    
    However, their serial order would be $txn_2 \rightarrow txn_1$, which violates the event order constraint (i.e., $txn_2$ will wrongly read the original value of $k1$). On the other hand, if $txn_2.ts < txn_1.ts$, then $txn_2$ will be successfully committed. However, $txn_1$ will be aborted as the writes will come too late. Aborting a transaction that represents an undo of an externally visible output or action may not be acceptable in \tsp applications. 
    
    Similarly, in other conventional CC protocols, either the results are in the wrong serial order or one of the transactions has to be aborted, which eventually will result in the wrong serial order upon a restart. In other words, conventional CC protocols are not yet ready for such \emph{event-driven} transaction execution.
\end{remark}

\begin{figure}[t]
    \centering
    \includegraphics[width=0.35\textwidth]{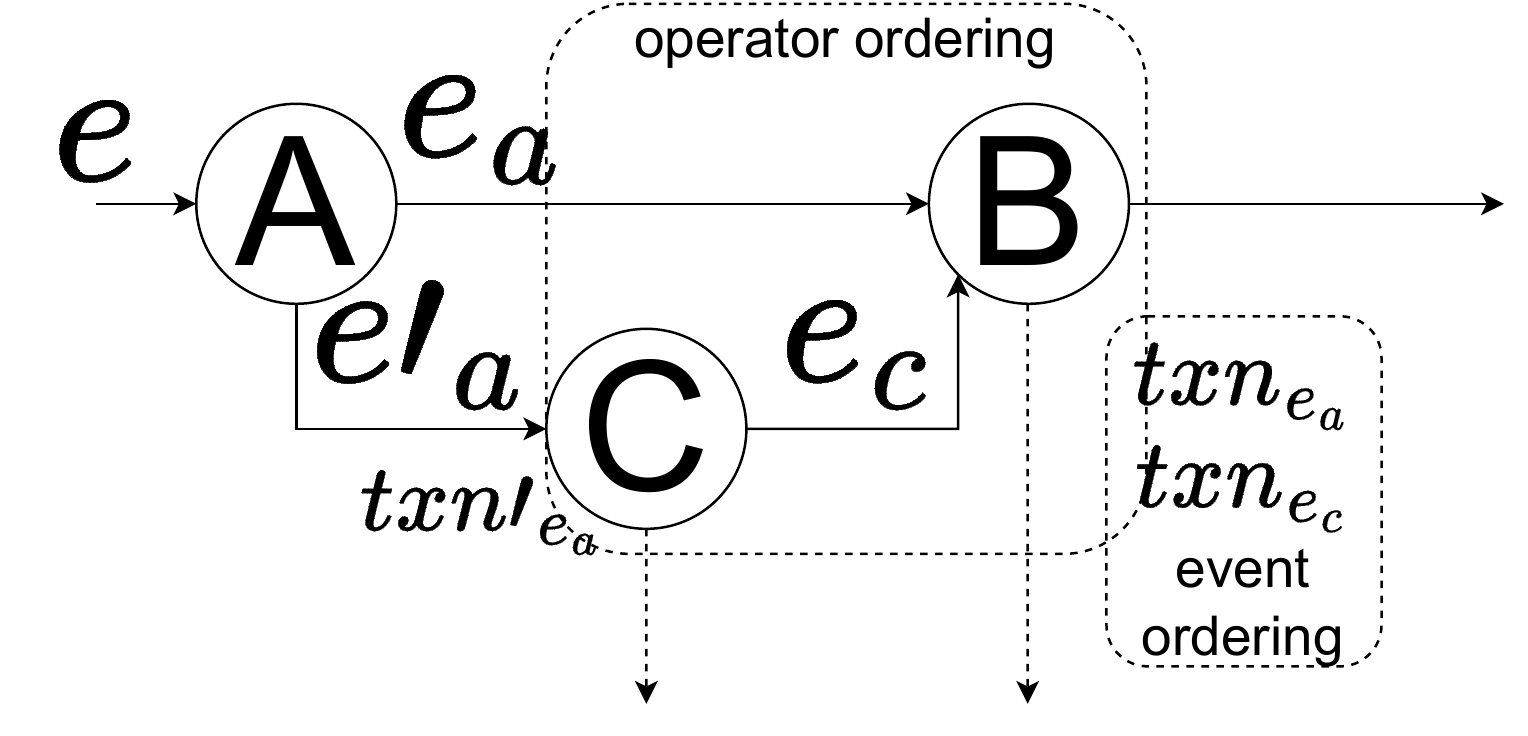}
    \caption{Example of timestamp assignment dilemma.}
    \label{fig:example_dilemma}
\end{figure}
\begin{remark}[Timestamp Assignment Dilemma]
\label{remark:dilemma}
\change{In TSP systems, a transaction triggered by a corresponding input event should ideally take effect with minimal latency after the event occurs~\cite{ACEP}. This concept is related to external consistency or linearizability in distributed systems theory.\margi{R1Q46}In this setting, the timestamp of the transaction is set to be the same as the timestamp of its corresponding triggering event. When all transactions are generated by external events, the transaction schedule is aligned with an external event sequence, satisfying the event ordering constraint.}

However, a dilemma arises when transactions can also be generated by internal events (i.e., outputs generated by operators) with timestamps assigned according to input events, as illustrated in Figure~\ref{fig:example_dilemma}. In the DAG, each operator can produce output streams and generate stream transactions. Suppose operator A receives input event $e$ and generates two events $e_a$ and $e_a'$, passed to operators B and C, respectively. Further, suppose operator C processes $e_a'$, generates $e_c$ as an output, and then passes $e_c$ to operator B. Now, let us assume operator B processes two events $e_a$ and $e_c$ and generates two transactions $txn_{e_a}$ and $txn_{e_c}$, respectively. In addition, operator C processes event $e_a'$ and generates transaction $txn_{e_a}'$. On the one hand, $txn_{e_a}$ must be committed jointly with $txn_{e_c}$, for otherwise, the event ordering constraint would be violated. On the other hand, $txn_{e_c}$ cannot be generated before $txn_{e_a}'$ is committed, for otherwise, this would violate the operator ordering constraint, which means $txn_{e_a}$ is unable to be committed. The system then runs into a deadlock situation, where $txn_{e_a}$ and $txn_{e_a}'$ are infinitely waiting for each other to be committed.

There are two approaches to prevent this dilemma: 1) an additional ordering constraint can be enforced between operators B and C, or 2) different timestamps could be assigned to the generated events: $e_a$, $e_a'$, and $e_c$. However, to date, we are not aware of any generally efficient solution to address such a dilemma.
\end{remark}

\subsection{Technologies Employed in \tsp Implementation}
\label{sec:implement}
This section delves into the practical aspects of implementing TSP systems, such as the choice of programming languages and APIs, system architectures, and the integration of various components to achieve a well-rounded TSP system. We also discuss performance metrics and evaluation criteria for TSP systems.

\subcompact
\subsubsection{Languages and APIs}
TSP systems should provide user-friendly and expressive languages that can easily define complex transactions, data manipulations, and processing logic over data streams while addressing aspects such as ordering properties, state management, and delivery guarantees. 
However, TSP systems do not yet have a standard transaction model or language, which complicates the selection of appropriate languages and APIs. Let us examine both declarative languages and functional languages in turn.

\paragraph{Declarative Languages:}
The STREAM system~\cite{motwani2002query,stream2003stream,Arasu:2006:CCQ:1146461.1146463} supports a declarative query language called CQL (Continuous Query Language), which is designed to handle both relational data and data streams. Coral8's~\cite{Coral8} continuous computation language (CCL) has a SQL-like syntax and supports both data streams and event streams. Franklin et al.~\cite{Franklin2009ContinuousAR} introduced TruSQL, a relational stream query language that fully integrates stream processing into SQL, including persistence.
These relational stream query\margi{R1Q52}languages are well-suited for TSP when handling a single query at a time. However, they may not sufficiently address TSP-specific issues like isolation and concurrent execution of queries. Declarative languages need to be extended to provide proper isolation guarantees. Moreover,  they need to facilitate the correct interaction among queries by incorporating features that explicitly express proper interactions or defining interaction constraints as separate rules.

\paragraph{Functional Languages:}
Functional languages~\cite{millwheel}, influenced by MapReduce-like APIs are more suitable for expressing state abstractions and complex application logic than SQL-like declarative languages. Streaming systems, like Flink embed functional/fluent APIs into general-purpose programming languages, which allow users to define custom dataflows akin to the Aurora system~\cite{Aurora}. This design is also present in TSP systems, such as FlowDBMS and TStream~\cite{AFFETTI202065,Affetti:2017:FIS:3093742.3093929,tstream}. These languages offer greater flexibility and expressiveness, enable developers to create custom processing logic, and leverage existing functional programming paradigms. However, they may introduce additional complexity and a steeper learning curve for developers new to TSP systems. 
Further research is needed to explore the trade-offs between the simplicity of declarative languages and the flexibility of functional languages, as well as to develop new programming constructs and abstractions that strike a balance between the two approaches, considering the transactional requirements of TSP systems.

\subcompact
\subsubsection{System Architectures}
\label{sec:architectures}
Given the properties and requirements discussed, TSP applications usually necessitate the use of SPEs in conjunction with data storage and analysis frameworks, such as database management systems (DBMSs), to create software architectures that integrate data storage, retrieval, and mining. Three approaches can be employed to construct a TSP system: 1) extending a DBMS, 2) embedding a DBMS in an SPE, and 3) composing a DBMS and an SPE. The choice of architectural approach for TSP systems depends on the specific requirements and constraints of the stream processing application. Each approach offers a unique set of trade-offs and considerations, making them more or less suitable for different use cases. By understanding the strengths and weaknesses of each approach, developers can make informed decisions when designing and implementing TSP systems to meet the needs of their applications. 
Next, we explore each approach in turn. 

\begin{figure*}
	\centering
	\begin{minipage}{\textwidth}
		\centering
		\subfloat[Extending a DBMS]{
			\includegraphics[width=0.32\textwidth]
			{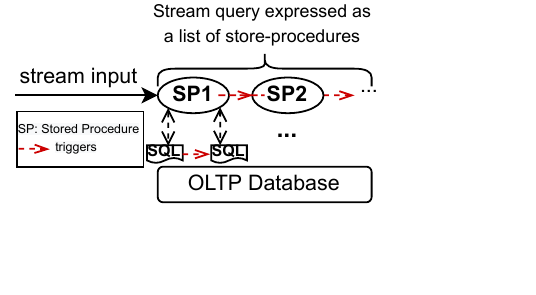}
			\label{fig:archit1}
		}\hfill
		\subfloat[Embedding a DBMS in a SPE
		]{
			\includegraphics[width=0.3\textwidth]
			{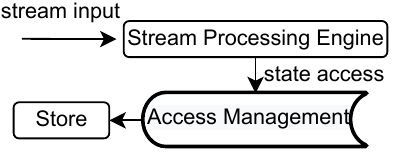}
			\label{fig:archit2}
		}\hfill   	  
		\subfloat[Composing a DBMS and an SPE]{
			\includegraphics[width=0.32\textwidth]
			{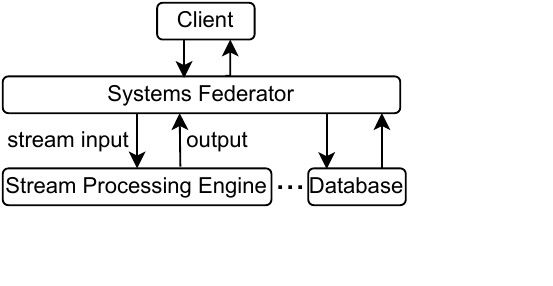}
			\label{fig:archit3}
		}   	                                    
		\caption{
			Illustrations of alternative system architectures of \tsp systems.
		}
		\label{figures:archit}
	\end{minipage}	
\end{figure*}

\paragraph{Extending a DBMS:}
Extending a DBMS for TSP involves incorporating stream processing functionality into traditional database management systems. By doing so, these systems aim to provide a unified platform for stream processing and traditional database management tasks while maintaining the strong consistency, isolation, and fault tolerance properties of traditional databases. \change{Since state is managed\margi{R1Q53}directly within the DBMS, it can be shared across queries and operators, and provide durability and consistency. Since transactions are supported in the underlying DBMS, it offers strong consistency guarantees. This setup is particularly advantageous in applications, such as financial systems, where the integrity and correctness of the data are paramount.} 

Notable examples of this approach include DataCell~\cite{Liarou2009DataCellBA}, MaxStream~\cite{Botan09}, Truviso Continuous Analytics system (TruCQ)~\cite{Sailesh10}, and S-Store~\cite{S-Store}. While extending a DBMS for TSP offers a unified platform, it also has limitations, such as difficulty in efficiently supporting native or hybrid stream processing applications and challenges in handling real-time processing and stateful operations. Consequently, alternative approaches might be more suitable for specific TSP applications.

Advantages of this approach include: a) leveraging the existing features and infrastructure of a DBMS for transactional support, query processing, and data management; b) simplifying the system architecture by integrating stream processing functionality within the DBMS, reducing the need for additional components or interfaces; and c) potentially providing strong consistency guarantees by directly utilizing the transactional mechanisms of the underlying DBMS. Disadvantages include: a) potentially being less flexible and adaptable to the specific requirements of stream processing applications, as it inherits the architectural constraints of the underlying DBMS; and b) possibly having limited scalability and performance due to the constraints of the underlying DBMS, which may not be designed for high-velocity, high-volume data streams.

\paragraph{Embedding a DBMS in an SPE:}
The embedding approach involves integrating an SPE with an embedded key-value store or DBMS to manage state, provide transactional support, and handle storage capabilities. This method enables stream processing systems to take advantage of the features of the embedded DBMS while maintaining the flexibility and efficiency of stream processing. \change{In this approach, state management occurs within the embedded DBMS, which can also be shared across queries and operators. However, the sharing model depends on the specific integration between the SPE and the embedded DBMS. Durability can be achieved via the underlying DBMS.}

Examples of this approach include Aurora~\cite{Aurora} and its successor Borealis~\cite{abadi2005design}, TStream~\cite{tstream} and its successor MorphStream~\cite{mao2023morphstream}. Embedding a DBMS within an SPE allows for various transactional models over streams but also introduces challenges such as integration complexity and potential performance bottlenecks or resource contention issues.
This approach is better suited for applications that require a balance between the performance and scalability of stream processing and the transactional support and data management capabilities of a DBMS. It can be particularly useful when the stream processing workload is dynamic and demands efficient state management and transactional support. Examples of such applications include real-time analytics, social network analysis, and large-scale data processing tasks like log analysis or clickstream processing.

Advantages of this approach include: a) enabling TSP systems to benefit from the features of both the SPE and the embedded DBMS, combining their strengths and providing a unified platform for stream processing; and b) offering improved performance and scalability by facilitating concurrent data processing and state management, as well as leveraging the distributed nature of modern SPEs.
Disadvantages of this approach include: a) necessitating careful integration of the embedded DBMS with the SPE, which may increase complexity and impact overall system latency; and b) the performance of the embedded DBMS might be influenced by the stream processing workload, potentially causing bottlenecks or resource contention issues. These problems can be mitigated by carefully tuning the embedded DBMS for the specific use case.

\paragraph{Composing a DBMS and an SPE:}
The composing approach involves using both a DBMS and an SPE in conjunction while keeping them as separate components. This method offers flexibility, adaptability, and optimizability for various application requirements. The SPE primarily focuses on processing streams, while the DBMS handles state management, transactional support, and storage capabilities. \change{In this approach, state is managed within the DBMS, while the SPE processes the streams. Sharing state across queries and operators depends on the communication and synchronization between the SPE and DBMS. Durability can be achieved via the underlying DBMS, but the communication latency  between the components can impact performance.}

Examples of this approach include Storage Manager for Streams (SMS)~\cite{SMS}. Although composing a DBMS and an SPE is more complex than other approaches and introduces additional challenges, such as performance overhead and latency in communication between the systems, it provides a flexible and scalable solution for building TSP systems that can leverage the strengths of both components.
The composing approach is well-suited for applications requiring high levels of flexibility, adaptability, and performance, as well as the ability to separately optimize and customize the SPE and the DBMS components. It is particularly useful for applications with diverse and evolving requirements, as it enables the system to be easily adapted or extended to accommodate new functionality or optimizations. Examples of such applications include IoT systems, sensor networks, and other data-intensive applications with varying processing and storage requirements.

Advantages of this approach include: a) offering the most flexibility and adaptability by allowing separate optimization and customization of the SPE and the DBMS, enabling tailored solutions for specific application requirements; and b) providing better scalability and performance by distributing workloads across multiple systems and enabling parallelization of processing tasks.
Disadvantages of this approach include: a) demanding additional effort to ensure integration correctness, which may increase development and maintenance complexity. This challenge can be mitigated by using standardized interfaces and middleware for communication between the SPE and the DBMS; and b) potentially exhibiting higher latency due to the need for communication and synchronization between the SPE and the DBMS. However, this can be mitigated by optimizing communication protocols and leveraging caching techniques to minimize data exchange overhead.

\subcompact
\subsubsection{Performance Metrics of TSP Systems}
In this subsubsection, we discuss the key performance metrics and evaluation criteria for transactional stream processing systems. These metrics help in understanding the efficiency and effectiveness of various TSP architectures, as well as in identifying the trade-offs associated with different system designs.

\paragraph{Throughput:}
Throughput is a measure of the number of transactions or events processed per unit of time. High throughput is desirable in TSP systems, as it indicates the system's capability to handle large volumes of data efficiently. Throughput can be affected by factors such as system architecture, resource allocation, and workload characteristics. TSP systems should be designed to maximize throughput while maintaining other performance guarantees, such as low latency and correctness.

\paragraph{Latency:}
Latency is the time taken for a transaction or event to be processed by the TSP system, from the moment it enters the system until it is completely processed. Low latency is crucial for TSP systems, as many real-time applications require timely processing of data. Latency is influenced by factors such as system architecture, data processing complexity, and resource utilization. TSP systems should be designed to minimize latency, ensuring that transactions are processed quickly without compromising other performance aspects.

\paragraph{Scalability:}
Scalability measures the ability of a TSP system to handle increasing amounts of data and concurrent transactions without significant performance degradation. As the volume of data and the number of concurrent transactions grow, TSP systems must be able to scale effectively to maintain low-latency access and fault tolerance. Techniques such as data sharding, partitioning, and replication can be employed to distribute the shared mutable state across multiple nodes, providing horizontal scalability.

\paragraph{Correctness Guarantee:}
Correctness guarantee in TSP systems refers to the extent to which a system can ensure data consistency and maintain the required transaction properties, as discussed in subsection~\ref{sec:properties}, in the presence of failures, network delays, or other factors. This includes support for various isolation levels, consistency models, and durability guarantees. TSP systems should be designed to provide the appropriate level of correctness guarantees based on the specific requirements of the stream processing application.

\begin{itemize}
    \item Isolation Levels: TSP systems should maintain proper isolation levels (e.g., serializable, snapshot isolation, read committed) to ensure that concurrent transactions do not interfere with each other and result in incorrect data processing.
    \item Consistency Models: Different TSP systems may adopt different consistency models (e.g., strong consistency, eventual consistency, or causal consistency) to provide a balance between data correctness and system performance. The choice of consistency model should be aligned with the application requirements and the tolerance for temporary inconsistencies.
    \item Durability Guarantees: TSP systems should ensure that once a transaction is committed, its effects are permanently recorded, even in the presence of failures. This can be achieved through techniques such as logging, checkpointing, and replication.
\end{itemize}

Balancing correctness guarantees with other performance metrics, such as throughput and latency, is crucial for achieving optimal performance in TSP systems. It is essential to recognize that the choice of the appropriate correctness guarantee is highly dependent on the application's requirements and the nature of the data being processed. For some applications, strong consistency and high isolation levels might be necessary, while for others, relaxed consistency models and lower isolation levels might be sufficient.

\compact
\section{Systems Offering Transactional Stream Processing}
\label{sec:systems}
We summarize six notable transactional stream processing (TSP) systems in Table~\ref{tab:comparison}. These TSP systems showcase the various approaches used to address the challenges in transactional stream processing. Each system offers unique features and each has made some design choices that cater to different application requirements. Let us explore each of these systems in turn, to gain some insight into the design and implementation of TSP systems.
\margi{R1Q7, R3Q5}
\begin{table*}[htbp]
\small
\centering
\caption{Key characterization of Six Transactional Stream Processing Systems}
\resizebox{\textwidth}{!}{%
\begin{tabular}{|P{1.8cm}|P{3cm}|P{4cm}|P{4cm}|P{4cm}|}
\hline
\textbf{System} & \textbf{Unique Features} & \textbf{Properties} & \textbf{Design Aspects} & \textbf{Implementation Details} \\
\hline
STREAM & first unified framework & ACID, ordering, at-least-once & unified transaction, lock-based approach & embedding a DBMS in an SPE \\
\hline
Botan et al. & pipelined architecture & ACID, ordering &  unified transaction, per-tuple transactions & embedding a DBMS in an SPE \\
\hline
S-Store & static partitioning & ACID, ordering, exactly-once, at-least-once & state-transaction, per-store-procedure-invocation transaction & extending a DBMS \\
\hline
Braun et al. & in-database stream analytics & ACID, ordering & state transaction, per-tuple transactions & extending a DBMS \\
\hline
FlowDB & transactional state management on Flink & ACID, ordering, exactly-once & state transaction, user-defined transaction boundary  & embedding a DBMS in an SPE \\
\hline
TStream /MorphStream & dual-mode scheduling & ACID, ordering, exactly-once & state transaction, per-tuple transactions & embedding a DBMS in an SPE \\
\hline
\end{tabular}
}
\label{tab:comparison}
\end{table*}

\paragraph{STREAM:}
STREAM~\cite{stream2003stream} is an early TSP system that introduces a unified framework for continuous query processing over data streams and relations. It supports ACID properties and offers at-least-once delivery guarantees. STREAM's design approach is based on a combination of a sliding-window model and a relational model for efficiently processing continuous queries. It employs a language called CQL (Continuous Query Language) to express queries, which can be translated into efficient query plans for execution. STREAM supports ACID properties and at-least-once delivery guarantees. The system focuses on ordering properties and state management properties by using windows and panes for efficient state handling. The architecture of STREAM is centered around a relational model, which simplifies the integration of stream processing and transaction processing.

\paragraph{Botan et al.:}
Botan et al.~\cite{botan2012transactional} proposed a TSP system that focuses on a pipelined architecture for efficient parallel execution. The system supports ACID properties, but does not explicitly specify the delivery guarantees. Its design approach is centered around per-tuple transactions, which ensures strong consistency guarantees and high isolation. Botan et al. also address the challenges of implementing transactional guarantees in a stream processing environment, such as handling failures and maintaining consistency among shared mutable state. They propose techniques for addressing these challenges, including state replication, checkpointing, and recovery mechanisms. The system adheres to ACID properties, while delivery guarantees are not explicitly mentioned. The system addresses fault tolerance and durability by employing state replication, checkpointing, and recovery mechanisms.

\paragraph{S-Store:}
S-Store~\cite{S-Store} is a state-transaction TSP system that builds upon H-Store, a distributed, main-memory relational database system. It extends H-Store with stream processing capabilities and supports ACID properties. S-Store offers various delivery guarantees, such as exactly-once and at-least-once processing semantics. The system employs a partitioning mechanism for efficient parallel execution of stream processing tasks. In S-Store, each transaction is represented as a stored procedure, which can be invoked by incoming data streams. These stored procedures manipulate shared mutable states within the context of ACID-compliant transactions, ensuring consistency and isolation among concurrent tasks.
S-Store supports ACID properties, exactly-once and at-least-once delivery guarantees, and employs a partitioning mechanism for distributed and parallelized execution. S-Store also focuses on fault tolerance techniques, including state replication, checkpointing, and recovery.

\paragraph{Braun et al.:}
Braun et al.~\cite{AIM15} presented a TSP system that focuses on in-database analytics by combining event-processing and real-time analytics within the same database. The system supports ACID properties. However, the delivery guarantees are not explicitly specified. Its design approach is centered around in-database stream processing, which allows the system to efficiently handle complex event processing and real-time analytics tasks without requiring external tools or components. The system supports ACID properties, while delivery guarantees are not specified. The system also emphasizes performance metrics and evaluation of TSP systems, showcasing its efficiency in handling large volumes of data and complex analytics tasks.

\paragraph{FlowDB:}
FlowDB~\cite{Affetti:2017:FIS:3093742.3093929} is a TSP system that integrates stream processing and consistent state management. It supports ACID properties and provides exactly-once delivery guarantees. The system's design approach is based on a state-transaction model, which allows for efficient management of shared mutable state. FlowDB features a language called FQL (Flow Query Language) for expressing continuous queries and supports the integration of custom stream processing operators and user-defined functions. FlowDB supports ACID properties, exactly-once delivery guarantees, and efficient state management. It also addresses fault tolerance and durability by employing checkpointing and recovery mechanisms.

\paragraph{TStream/MorphStream:}
TStream~\cite{tstream} and its successor MorphStream~\cite{mao2023morphstream} are TSP systems that emphasize efficient concurrent state access on multicore processors. These systems support ACID properties and offer exactly-once delivery guarantees. TStream's design approach includes a unique dual-mode scheduling strategy that combines transactional and parallel modes, to enable the system to maximize parallelism opportunities offered by modern multicore architectures. TStream's dynamic restructuring execution strategy further improves concurrency by adapting the execution plan based on observed state access patterns during runtime. Unlike other works, TStream focuses on improving the performance of concurrent state access on multicores with efficient execution mechanisms, and delivery guarantees.
\compact
\color{black}
\section{Applications/Scenarios Leveraging TSP}
\label{sec:app}
\tsp arises in varying domains, such as healthcare~\cite{ACEP}, the Internet of Things (IoT)~\cite{botan2012transactional}, and e-commerce~\cite{Affetti:2017:FIS:3093742.3093929}). 
Table~\ref{tab:scenario_summary} summarizes thirteen scenarios. Each application encompasses diverse features, transactional models, and implementations of \tsp systems. We can categorise them into four scenarios: \emph{stream processing optimization}, \emph{concurrent stateful processing}, \emph{stream \& DBMS integration}, and \emph{recoverable stream processing}.

\begin{table*}[ht]
\centering
\small
\begin{tabular}{|l|l|l|l|l|l|}
\hline
 & \textbf{Scenario} & \textbf{Ordering} & \textbf{ACID} & \textbf{State Management} & \textbf{Reliability} \\
\hline
\multicolumn{6}{|l|}{\textbf{Stream Processing Optimization}} \\
\hline
1 & Sharing Intermediate Results & event, operation & ACID & read-write, inter-operator & consistency \\
\hline
2 & Multi-Query Optimization & event & ACID & read-write, global & delivery guarantee \\
\hline
3 & Deterministic Stream Operations & event & ACID & read-write, intra \& inter-operator & - \\
\hline
4 & Prioritizing Query Scheduling & event & ACID & read-write, inter-query & delivery guarantee \\
\hline
\multicolumn{6}{|l|}{\textbf{Concurrent Stateful Processing}} \\
\hline
5 & Ad-hoc Queryable States & event, operation & ACID & read-write, global & consistency, durability \\
\hline
6 & Concurrent State Access & event, operation & ACID & read-write, intra \& inter-operator & consistency, durability \\
\hline
7 & Active CEP & event & ACID & read-write, inter-Query & delivery guarantee, durability \\
\hline
\multicolumn{6}{|l|}{\textbf{Stream and DBMS Integration}} \\
\hline
8 & Streaming Ingestion & event & ACID & read-write, global & delivery guarantee, durability \\
\hline
9 & Streaming OLTP & operation & ACID & read-write, per-transaction & consistency, durability \\
\hline
10 & Streaming OLAP & event (optional) & snapshot & read-write, global & delivery guarantee, fault tolerance \\
\hline
\multicolumn{6}{|l|}{\textbf{Robust Stream Processing}} \\
\hline
11 & Shared Persistent Storage & operation & ACID & read-write, global & consistency, durability \\
\hline
12 & Transaction Identifier & operation & ACID & read-write, global & consistency, durability \\
\hline
13 & Fault Tolerance Outsourcing & operation & ACID & read-write, global & fault tolerance, durability \\
\hline
\end{tabular}
\caption{Summary of application scenarios and their demands of TSP properties.}
\label{tab:scenario_summary}
\end{table*}

\subcompact
\subsection{Stream Processing Optimization}
Several works have proposed the consistent management of shared mutable states to optimize stream processing. Below we discuss these works in the context of stream processing optimization across four use cases: \emph{sharing intermediate results}, \emph{multi-query optimization}, \emph{deterministic stream operations}, and \emph{prioritizing query scheduling}.

\textbf{Sharing Intermediate Results.}
In the STREAM system~\cite{stream2003stream}, nearly identical states, or synopses, within a query plan are kept in a single store to reduce storage redundancy. Operators access their states exclusively via a stub interface. As operators are scheduled independently, they require slightly different data views, so STREAM employs a timestamp-based execution mechanism for correctness. Ordering properties are crucial in this scenario. Event ordering constraints maintain the data stream order, while operation ordering constraints ensure transaction operations' order. ACID properties maintain the correctness and consistency of shared mutable states. State management properties require read-write state types and inter-operator access scope for managing shared states among different operators.

\textbf{Multi-Query Optimization.}
Ray et al.~\cite{SPASS} introduced the SPASS (Scalable Pattern Sharing on Event Streams) framework, which optimizes time-based event correlations among queries and shares processing effectively. The optimizer identifies a shared pattern plan, maintaining an optimality bound. The runtime then uses shared continuous sliding view technology for executing the shared pattern plan. A sequence transaction model on shared views defines the correctness of concurrent shared pattern execution. SPASS doesn't modify existing states but selects, inserts, and deletes shared states like sliding views. Ordering properties are crucial for maintaining the correct order of pattern queries. Event ordering constraint preserves data stream order, while operation ordering constraint is less important as sharing is among queries. ACID properties maintain consistency and correctness of shared states like sliding views. State management properties need read-write state types and global access scope for managing shared states among pattern queries. However, some implementation details are not specified in the original paper (e.g., sliding views' data layout, key used for searching shared sliding views).

\textbf{Deterministic Stream Operations.} Handling out-of-order streams is often a performance bottleneck due to the conflict between data parallelism and order-sensitive processing. While data parallelism improves throughput by processing events concurrently, it can cause events to be handled out-of-order. Most solutions use locks or non-lock algorithms like sorting~\cite{survey_hwstream}. Brito et al.~\cite{OOO_STM} proposed an interesting non-lock approach using software transactional memory (STM) for stream processing. They model processing a batch of input data at order-sensitive operators as a transaction and pre-assign commit timestamps, effectively imposing order. Events received out-of-order or conflicting are processed in parallel optimistically but aren't output until all preceding events are completed, ensuring consistent operator states. Ordering Properties are crucial for managing out-of-order streams, with event ordering constraint needed to maintain data stream order. Operation ordering constraint is less relevant. ACID Properties maintain correctness and consistency of shared mutable states, especially when handling out-of-order and conflicting events. State Management Properties require read-write state types and intra- and inter-operator access scope for managing shared states within and among different operators.

\textbf{Prioritizing Query Scheduling.} Handling potentially infinite data streams requires continuous queries with window constraints to limit tuple processing. Most implementations execute sliding window queries and window updates serially, implicitly assuming a window cannot be advanced while accessed by a query. Golab et al.\cite{golab2} argue that concurrent processing of queries (reads) and window-updates (writes) is necessary for prioritized query scheduling to improve answer freshness. They model window updates and queries as transactions with atomic sub-window reads and writes, which can lead to read-write conflicts. Golab et al.\cite{golab2} prove traditional conflict serializability is insufficient and define stronger isolation levels restricting allowed serialization orders following event ordering. Ordering Properties are crucial for prioritizing query scheduling, with event ordering constraint ensuring correct data stream order. ACID Properties maintain correctness and consistency of shared mutable states when handling concurrent reads and writes. State Management Properties require read-write state types and inter-query access scope for managing shared states among different queries and window updates.

\subcompact
\subsection{Concurrent Stateful Processing}
In this scenario, application workloads consist of both ad-hoc and continuous queries, which may access and modify common application states for future reference~\cite{flinkstate}. We will discuss three representative applications in the context of concurrent stateful processing: \emph{ad-hoc queryable states}, \emph{concurrent state access}, and \emph{active complex event processing}, focusing on the properties of TSP required by each application.

\textbf{Ad-hoc Queryable States.} Ad-hoc queries, or snapshot queries, can be submitted to an SPE anytime, executed once, and provide insights into the system's current state. They may be used to obtain further details in response to continuous query result changes. In Botan et al.'s example~\cite{botan2012transactional}, real-time sensors generate temperature measurements to ensure temperature-sensitive devices operate within design specifications. When a temperature reading falls out of the operating range, it triggers an alert. The SPE must ensure table updates and stream temperature readings are executed in the correct order, demanding event and operation ordering constraints. To maintain the specifications table's integrity and prevent data corruption, ACID Properties are necessary. The state is read-write, as the table needs updates, and the access scope is global, as all incoming temperature readings need to probe the table. Delivery Guarantee and Fault Tolerance are vital for ensuring accurate temperature reading processing and system recovery from potential failures.

\textbf{Concurrent State Access.}
In applications like Ververica Streaming Ledger~\cite{Transactions2018} (SL), operators like parser, deposit, transfer, and sink may need to share access to states, such as account and asset data. To process transactions correctly and maintain data consistency, event and operation ordering constraints are crucial. ACID Properties are necessary to ensure shared state accuracy and prevent data corruption during concurrent access. The state is read-write, as the account and asset data need updates, and the access scope includes intra and inter-operator, as multiple operators and their replica instances access and modify the shared stat. Delivery Guarantee, Fault Tolerance, and Durability are critical for maintaining accurate and consistent transaction processing. CAP Theorem considerations must be taken into account when designing the system to balance consistency, availability, and partition tolerance.

\textbf{Active Complex Event Processing.}
In the realm of stream processing, active complex event processing is a method that continuously monitors and analyzes a series of real-time events to detect certain patterns or sequences. Wang et al.~\cite{ACEP} have applied this concept to a hospital infection control application.
The system uses sensor devices to generate real-time data on healthcare workers' (HCWs) behaviors such as ``exit'', ``sanitize'', and ``enter''. These events are then analyzed by pattern queries that aim to detect any violations of hospital hygiene rules. The status of all HCWs, whether static or dynamic, is stored in tables. In this application, it is crucial to maintain the correct order of events and handle concurrent accesses and updates during stream execution. This requires event ordering constraint and ACID Properties to ensure data consistency and prevent conflicts. The state management in this scenario involves read-write states with inter-query access scope, as multiple pattern queries may read or update the tables concurrently. Reliability and delivery guarantee properties, such as delivery guarantee, fault tolerance, and durability, are essential for accurately and consistently detecting violations of hospital hygiene rules.

\subcompact
\subsection{Stream and DBMS Integration}
The integration of SPEs with DBMSs is becoming increasingly important~\cite{Tatbul10}. Scenarios such as stream data ingestion (i.e., \emph{Streaming Ingestion}), implementing OLTP queries in alternative ways (\emph{Streaming OLTP}), and mixed stream and analytic queries (\emph{Streaming OLAP}) can be well supported by TSP systems. In the following sections, we discuss each of these scenarios and their requirements concerning TSP properties.

\textbf{Streaming Ingestion.} Streaming ingestion is an essential process for organizations handling large volumes of data, as it enables more timely access to incremental results compared to traditional batch ingestion. This approach processes smaller microbatches throughout the day, which requires proper management of ordering and state properties. A notable example by Meehan et al.\cite{meehan2017data} involves self-driving vehicles, where the value of sensor data decreases over time. In this case, timely processing and storage of time series data are critical for making valuable decisions. The authors adapt TPC-DI\cite{tpc_di}, a standard benchmark for data ingestion, to assess streaming ingestion effectiveness while considering new data dependencies introduced by breaking large batches into smaller ones. In this scenario, maintaining the correct order of time series data is crucial, necessitating event ordering constraint. ACID properties are required to ensure the correctness and consistency of ingested data, especially with high sample rates and large data volumes. Read-write state types are needed for efficient management and persistence of ingested data. Delivery guarantee is essential for processing and persisting ingested data accurately and consistently in near real-time. Additionally, durability is necessary to ensure data availability for analysis and decision-making, even after a system failure or outage.

\textbf{Streaming OLTP.} Streaming OLTP addresses traditional OLTP workloads using streaming queries. Chen and Migliavacca~\cite{chen2018streamdb} propose \emph{StreamDB}, a TSP-based system. Streaming OLTP requires ACID Properties, State Management Properties (e.g., database partitioning), and fault tolerance. \emph{StreamDB} uses three operators in a streaming query: 1) \emph{Source} operator, receiving transactions and sending them to downstream data operators; 2) \emph{Data} operator, managing a portion of a database, executing transactions, and producing results; 3) \emph{Sink} operator, receiving transaction responses. \emph{StreamDB} reduces lock contention by distributing the database among multiple data operators. However, creating an optimal stream dataflow graph for diverse OLTP workloads remains an open question. In this scenario, operation ordering constraint is crucial for correct transaction processing and minimizing lock contention. ACID properties ensure database correctness and consistency during transaction processing. Read-write state types are required for managing the database and transactions, while maintaining consistency. Consistency, Durability, and Fault Tolerance are vital for accurate and consistent transaction processing and for maintaining database consistency even after system failures or outages.

\textbf{Streaming OLAP.}
Organizations often need real-time analysis of data streams for immediate decision-making, and several related applications have been described in the literature~\cite{AIM15,streamingcube,Gtze2019SnapshotIF}. The Huawei-AIM workload~\cite{AIM15} features a three-tier architecture with storage, an SPE, and real-time analytics (RTA) nodes. RTA nodes push analytical queries to storage nodes, merge partial results, and deliver final results to clients. To meet the service level objective (SLO), a consistent state (or snapshot) must not be older than a certain bound. In this scenario, event ordering constraint ensures correct analytical query processing. ACID properties can be relaxed (e.g., snapshot isolation) to maintain data correctness and consistency. Read-write state types manage the database and ensure the consistent state meets the SLO. Delivery Guarantee, Consistency, Durability, and Fault Tolerance ensure accurate query processing, data persistence, and system recovery from failures.

\subcompact
\subsection{Robust Stream Processing}
In addition to requirements such as \emph{scalability} and \emph{low latency}, many critical streaming applications demand SPEs to recover quickly from failures~\cite{requirements}. Consequently, considerable effort has been dedicated to achieving fault tolerance in SPEs. We discuss previous attempts to employ transaction-like concepts to ensure high availability and fault tolerance in stream processing, focusing on their demands for properties of TSP discussed in Section~\ref{sec:properties}.

\textbf{Shared Persistent Storage.}
MillWheel~\cite{millwheel} uses an event-driven API for stateful computations and stores input and output data persistently. It relies on remote storage systems like BigTable~\cite{bigtable} for managing state updates and handling fault tolerance through data replication. This relates to Durability in ACID properties and Reliability and Delivery Guarantee properties. MillWheel ensures Atomicity by encapsulating all per-key updates in a single atomic operation. While it enforces operation ordering constraints, strict event ordering constraints are not provided. State Management Properties are vital in MillWheel due to its dependence on shared persistent storage.

\textbf{Transaction Identifier.}
Trident's ``transactional topology''~\cite{trident} processes small batches of tuples as a single operation and assigns unique transaction identifiers (\emph{TXID}), relating to Ordering Properties. TXID, logged in external storage along with operator state, addresses Atomicity and Durability in ACID properties. If TXID mismatch occurs, a batch must be resubmitted, requiring strict transaction processing ordering (operation ordering constraint) and potentially limiting throughput. State Management Properties are essential in this approach, as it depends on read-write states. Trident provides strong Delivery Guarantee and Fault Tolerance, but may lose intermediate results during failures due to disregarding buffered input states.

\textbf{Fault Tolerance Outsourcing.}
Ishikawa et al.~\cite{BDA18} propose integrating fault tolerance into an OLTP engine for data stream processing. This involves backing up data streams in an in-memory database system instead of a file system, addressing Durability in ACID properties and Reliability and Delivery Guarantee properties. It enforces operation ordering constraints, similar to H-Store's state partitioning transaction processing. However, outsourcing fault tolerance can strain the remote store during data spikes, potentially impacting other applications sharing the store. This approach also requires CAP Theorem considerations, as the choice of an in-memory database system might involve trade-offs between consistency and availability.


\begin{remark}[\tsp is not just nice to have, but sometimes a must]
\change{
By employing modern SPEs, existing workarounds, such as using external databases to store shared application states, can lead to significant additional programming effort~\cite{ACEP}, poor performance~\cite{S-Store}, and even incorrect results~\cite{golab2,S-Store-demo}. 
This problem is exacerbated if more complex shared mutable state storage and retrieval queries, such as range look-ups are further required. In contrast, \tsp systematically manages concurrent accesses to shared application states with transactional correctness. This even leads \tsp to have the potential to better support traditional database workloads (i.e., OLTP and OLAP) as well. 
}
\end{remark}

\begin{remark}[\tsp-based applications have diverse requirements]
Some \tsp-based applications do not require insertion and deletion operations at all, while others do not need to update shared mutable states. Also noteworthy is that transactional dependency is rare among applications, which means that in most cases, the input parameters in a transaction are predetermined from the triggering events. Targeting a narrowed application domain, a \tsp system can take advantage of these diverse requirements to simplify its design and improve system performance. To date, there is no standard benchmark for \tsp  systems~\cite{Tatbul2018}, which must include comprehensive performance metrics, diverse workload features, and meaningful application scenarios.
The applications that we list in Table~\ref{tab:scenario_summary} may serve as a starting point for the construction of a standard benchmark. \change{However, more\margi{R2Q3}applications may need to be included, arranged according to their particular application's features.}
\end{remark}
\compact
\section{Research Outlook}
\label{sec:future}
In this section, we offer a perspective on future research directions of TSP. 

\subcompact
\subsection{Novel Applications}
The rise of IoT generates real-time data that needs immediate processing. Traditional big data applications were designed for large static datasets, but modern applications demand more. We foresee novel streaming applications benefiting from \tsp solutions as the range of applications served by SPEs widens. Current research, like NebulaStream~\cite{nebulastream}, explores systems meeting these requirements. We discuss various application areas, including online machine learning/stream mining, mixed batch/stream transactional workloads, streaming materialized views, and cloud applications.

\textbf{Optimization for Stateful Stream Processing.}
Shahvarani and Jacobsen's IBWJ~\cite{IBWJ} accelerates sliding window joins by using a shared index data structure, reducing redundant memory access and improving performance. As new tuples arrive, the index structure is updated, raising concurrency control issues. Shahvarani et al.\cite{IBWJ} propose a low-cost concurrency control mechanism for high-rate update queries. A \tsp system could naturally handle this concurrency problem, providing durability when required. Despite its potential, we are unaware of any practical implementation of this approach.

\textbf{Online Machine Learning/Stream Mining.}
The rising demand for data stream analysis necessitates online learning and mining. Current efforts support continuous queries (CQs) referencing non-streaming resources like databases and ML models~\cite{DSPCC}. Model-based streaming systems, such as anomaly detectors, require regular model updates without significantly increasing operational costs~\cite{millwheel}. Due to the lack of transactional support in traditional SPEs, implementing emerging streaming learning and mining algorithms can be challenging~\cite{10.1007/978-3-030-19274-7_10}.
\change{Although existing batch-based ML training (like TensorFlow) may not need to care for inconsistencies in the state they handle, a streaming ML scenario may prohibit such inconsistencies as each input data may be allowed to be used, up to a limited threshold, and any inconsistency may lead to significantly lower training quality.}\margi{R1Q67}It thus remains an interesting future work to study how those novel training and mining use cases can be supported efficiently in \tsp  systems, which bring features, such as elastic scaling, fault tolerance guarantees, and shared state consistency to users, even at the virtual space~\cite{ooi2022sense}.

\textbf{Mixed Batch/Stream Transactional Workloads.}
Many enterprise applications, particularly in finance and IoT, generate mixed workloads with continuous stream processing, OLTP, and OLAP. DeltaLake~\cite{10.14778/3415478.3415560} allows streaming jobs to write small objects into a table with low latency and coalesce them into larger objects later. Fast ``tailing'' reads are also supported for treating a Delta table as a message bus. Tatbul~\cite{Tatbul10} outlines challenges in streaming data integration, including common semantic models, optimization, and transactional issues. These challenges persist due to diverse applications and systems focusing on limited feature sets.

\textbf{Streaming Materialized Views.}
Traditional materialized views (MVs) are not optimized for high-velocity data stream processing, leading to the need for streaming materialized views (SMVs). SMVs must handle high-velocity inputs, update states with random access patterns, and share updated states among concurrent entities. Recent works, such as S-Query~\cite{verheijde2022s} and Umbra's continuous view scheme~\cite{winter13meet}, have proposed solutions, but the former lacks strict ACID guarantees, and the latter has yet to be compared with state-of-the-art transactional stream processing systems like S-Store~\cite{S-Store}, TStream~\cite{tstream}, and TSpoon~\cite{AFFETTI202065}. The distinction in use cases drives a clear separation of concerns and further investigation into optimizing SMVs.

\textbf{Cloud Applications.}
\margi{R2Q4}
\change{Existing SPEs often lack transactional facilities needed for cloud applications that require advanced business logic and coordination~\cite{evolution20}.} One example is stateful function-as-a-service, which demands ACID transactions, global state consolidation, and debugging and auditing capabilities~\cite{katsifodimos2019operational, akhter2019stateful}. These requirements are similar to those of transactional stream processing (TSP). However, it's unclear if TSP systems like S-Store~\cite{S-Store} can fully satisfy these requirements. Further exploration is needed, including supporting debugging~\cite{stream_debug} and isolation~\cite{254432} in stateful stream processing as microservices.

\subcompact
\subsection{Novel Hardware Platforms}
Modern hardware advancements have made servers with hundreds of cores and several terabytes of main memory available.
Such advancements have driven researchers to rethink \tsp  systems and put emerging hardware platforms to good use~\cite{survey_hwstream}. 
Next, we take a closer look at multi-/many-core architectures, non-volatile storage, and trusted computing platforms.

\textbf{Multi-/Many-core Architectures.} 
Supporting shared mutable states in \tsp systems can create bottlenecks due to concurrent state accesses. TStream~\cite{tstream} is a recent example that effectively utilizes multicore CPUs to improve concurrent shared state access performance through dual-mode scheduling and a dynamic transaction restructuring mechanism. However, current \tsp systems still face scalability challenges with complex workloads and input dependencies. Further research is needed to enhance \tsp systems for complex workloads, emerging multi-/many-core architectures with high-bandwidth memory, and multi-node settings while maintaining correctness guarantees~\cite{Tatbul2018}.

\textbf{Non-Volatile Storage.}
Non-Volatile Memory (NVM) is an emerging technology offering byte-addressability and low latency of DRAM along with persistence and density of block-based storage media, but with limited cell endurance and read-write latency asymmetry. Fernando et al.\cite{NVM} explored efficient approaches for analytical workloads on NVM, potentially laying the foundation for future \tsp systems\cite{TSM_NVM}. NVMe-based SSDs can deliver high performance in terms of latency and peak bandwidth. Lee et al.~\cite{SSD_Stream} investigated performance limitations of SPEs managing application states on SSDs, showing query-aware optimization can improve stateful stream processing on SSDs. Their work is valuable for \tsp systems with strict \acid and streaming property requirements, but more research is needed.

\textbf{Trusted Computing Platforms.}
The need for low latency and local processing of sensitive IoT data calls for edge stream processing. However, edge devices are vulnerable to attacks due to limited power and computing capacity, posing severe security threats to sensitive data. A potential solution~\cite{streamboxtz} is trusted computing platforms (TCPs), which protect data and code within isolated, encrypted memory areas. Bringing \tsp to TCPs is nontrivial and requires further research, particularly in handling limited memory for transactional stateful stream processing~\cite{a15060183}. Additionally, scaling systems to multiple TCPs in a distributed environment presents challenges due to each computing node's computational limits.
\compact
\section{Conclusion}
\label{sec:conclusion}
\margi{R1Q5}
\change{
In this survey, we provided a comprehensive overview of Transactional Stream Processing (TSP), and addressed key concepts, techniques, and challenges to be overcome, in order to ensure reliable and consistent data stream processing. We introduced terms, definitions, and a conceptual framework for TSP systems and presented a taxonomy that offers a structured understanding of various approaches and models for integrating transactional properties with streaming requirements. We also discussed several notable TSP systems, each showcasing unique features and design choices that were made, to cater to different application requirements. These  systems offer insight and inform designers about the alternative choices they will need to make when designing and implementing a novel TSP system. We highlighted various TSP applications and use cases, such as stream processing optimization, concurrent stateful processing, and stream and DBMS integration. Finally, we explored some open challenges and suggest future directions for TSP research and development, including novel applications and hardware platforms. This survey serves as a valuable resource for researchers and practitioners. It aims to inspire others to pursue work in this field and develop efficient, reliable, and scalable TSP systems for diverse application domains.
}
\compact
\section{Declarations}
The authors have no financial or proprietary interests in any material discussed in this article.

\begin{acknowledgements}
This work is supported by the National Research Foundation, Singapore and Infocomm Media Development Authority under its Future Communications Research \& Development Programme (FCP-SUTD-RG-2021-005), the SUTD Start-up Research Grant (SRT3IS21164), the DFG Priority Program (MA4662-5), the German Federal Ministry of Education and Research (BMBF) under grants 01IS18025A (BBDC - Berlin Big Data Center) and 01IS18037A (BIFOLD - Berlin Institute for the Foundations of Learning and Data). Shuhao Zhang'work is partially done while working as a Postdoc at TU Berlin.
\end{acknowledgements}

\bibliographystyle{spmpsci}
\bibliography{mybib}

\end{document}